%% file: ijdl.tex
\RequirePackage{fix-cm}
\documentclass[twocolumn]{svjour3}          
\smartqed  
\usepackage{graphicx}
 \usepackage{todonotes}
 \usepackage[numbers]{natbib}
\usepackage{multirow}
\usepackage{subfig}
\usepackage[section]{placeins}
\usepackage{enumitem}
\usepackage{comment}
\usepackage{array}
\usepackage{amsmath}
\usepackage{gensymb}
\newcolumntype{P}[1]{>{\centering\arraybackslash}p{#1}}

\begin{document}

\title{Investigating Exploratory Search Activities based on the Stratagem Level in Digital Libraries} 
	


\author{Zeljko Carevic    \and         Maria Lusky     \and         Wilko van Hoek  \and         Philipp Mayr 
}


\institute{ Zeljko Carevic  \and  Wilko van Hoek \and Philipp Mayr\\
			  \at GESIS –- Leibniz Institute for the Social Sciences\\
              Unter Sachsenhausen 6-8\\
              50667 Cologne, Germany\\
              \email{zeljko.carevic@gesis.org} \\
              \and Maria Lusky \\
              \at RheinMain University of Applied Sciences \\
	          Unter den Eichen 5\\
	          65195 Wiesbaden, Germany \\    
              \email{maria.lusky@hs-rm.de} 
}

\date{Received: date / Accepted: date}

\maketitle

\begin{abstract} 
In this paper we present the results of a user study on exploratory search activities in a social science digital library. We conducted a user study with 32 participants with a social sciences background -- 16 postdoctoral researchers and 16 students -- who were asked to solve a task on searching related work to a given topic. The exploratory search task was performed in a 10-minutes time slot. The use of certain search activities is measured and compared to gaze data recorded with an eye tracking device. We use a novel tree graph representation to visualise the users' search patterns and introduce a way to combine multiple search session trees. 
The tree graph representation is capable to create one single tree for multiple users and to identify common search patterns. In addition, the information behaviour of students and postdoctoral researchers is being compared. 
The results show that search activities on the stratagem level are frequently utilised by both user groups. The most heavily used search activities were keyword search, followed by browsing through references and citations, and author searching. The eye tracking results showed an intense examination of documents metadata, especially on the level of citations and references. When comparing the group of students and postdoctoral researchers we found significant differences regarding gaze data on the area of the journal name of the seed document. 
In general, we found a tendency of the postdoctoral researchers to examine the metadata records more intensively with regards to dwell time and the number of fixations. 
By creating combined session trees and deriving subtrees from those, we were able to identify common patterns like economic (explorative) and exhaustive (navigational) behaviour. Our results show that participants utilised multiple search strategies starting from the seed document, which means, that they examined different paths to find related publications.

\end{abstract}

\input{intro.tex}

\input{relatedwork.tex}

\input{userStudy.tex}
\input{methodology.tex}

\input{demographics.tex}

\section{Results}\label{results}

\input{usagefrequency.tex}

\input{eyetracker.tex}

\input{graph.tex}

\input{diversity.tex}

\input{discussion.tex}

\input{conclusion.tex}

\input{acknowledgement.tex}


%
%


\bibliographystyle{spbasic}      


\bibliography{ijdl}

\end{document}

%% file: intro.tex
\section{Introduction}\label{intro}
Digital Libraries (DLs) offer direct access to a vast number of bibliographic records and full texts. The amount of content a user needs to assess becomes difficult to manage which is often referred to as information overload and typically leads to highly fragmented retrieval sessions in which users perform various types of search activities \cite{ODay/Jeffries:93}. Search activities of users in scholarly information systems are exploratory or focused on known items \cite{marchionini2006exploratory}. \\
During the past, different models have been proposed that aim to model the information seeking behaviour, e.g. the berrypicking model by Bates \cite{bates1989design}. 
Based on empirical studies of the information seeking behaviour of experienced library users Bates identified four levels of search activities that, amongst others, differ in their complexity: \textit{moves}, \textit{tactics}, \textit{stratagems} and \textit{strategies}. A move is the lowest unit of search activities like entering a query term or selecting a certain document. Tactics are described as a combination of several moves like the selection of a broader search term or breaking down complex search queries into subproblems. Bates defines a stratagem as follows: "a stratagem is a complex of a number of moves and/or tactics, and generally involves both a particular identified information search domain anticipated to be productive by the searcher, and a mode of tackling the particular file organization of that domain" \cite{bates1990should}.\\
Hence, a stratagem could be for instance a "journal run" where a user identifies a journal to be productive for his or her research and browses the latest publications in that journal. Another example for a stratagem is to follow references in a certain seed document that might lead to potentially relevant material. Finally, strategies are combinations of moves, tactics and stratagems, thus, forming the highest search activity as they cover the whole information seeking process.

Although modern DLs widely support various search activities, the attention by the scientific community on their empirical evaluation is comparatively little. In \cite{carevic2016survey} we present a first approach on gathering a deeper understanding on the usage of stratagems by conducting an online survey with 128 respondents from twelve different fields of research. The results of the survey provided us with insights regarding two aspects of stratagems: 
\begin{itemize}
\item{Usefulness}: The results of the survey underline the usefulness of stratagems, in particular, citations, references and journal / conference runs. When looking for related content to a given document, the respondents preferred to utilise references, citations and keywords.
\item{Experience}: We created different clusters of respondents to look for significant differences which were found for the groups of \em Senior Researchers (postdoctoral researchers, faculty and professors) \em and \em Students\em.  
The former considered "citation chaining" as a more valuable stratagem to look for related documents for a given topic whereas the latter preferred to use keywords. 
\end{itemize}

Although our online survey \cite{carevic2016survey} provided us with valuable insights on the general usage of stratagems, we are missing a qualitative perspective where we gather a deeper understanding on the intended use of certain search activities. For this reason, we conducted a user study with 32 participants from the social sciences domain who were invited to our user lab and asked to solve a given search task. The participants were provided with a topic on \em educational inequality \em and a seed document that was relevant to the given search task. The exploratory task was then to look for content that is similar to the given seed document in a very limited time slot. Furthermore, we observed the participants during the task and recoded their gaze data with an eye tracking device. With this approach, we aim to gain insights on the perceived relevance of certain search activities. In order to verify the results of the online survey we recruited participants from two levels of experience. The first 16 participants were students while the remaining 16 participants were postdoctoral researchers from the social sciences.  
\newline

To date, numerous studies have been presented that investigate exploratory search and search activities from different perspectives. Main differences between the studies can usually be found in the study context, e.g. \linebreak transaction-log studies or lab studies, the environment (web search, information portal search) and the methodology to investigate users' search behaviour. The methodologies range from descriptive counts and user feedback (e.g. \cite{fields2005designing},\cite{wilson2008longitudinal}), qualitative feedback and interviews (e.g. \cite{athukorala2013information}), gaze data (e.g. \cite{kules2009exploratory}) to mixed-methods (e.g. \cite{wildemuth2004effects}). 

Present studies on exploratory search focus on different aspects like query reformulation \cite{fields2005designing}, measuring the perceived relevance of result lists, facets and queries using eye tracking software \cite{kules2009exploratory}, maximal repeating pattern (MRP) of state transitions between experts and students \cite{wildemuth2004effects} or the usage of exploratory features across different phases of the information seeking process in comparison to keyword searches \cite{wilson2008longitudinal}.
In our study we differ from the present literature on two levels: a) the scope of our study where we adopt a use case which describes a specific user behaviour in a real life DL and b) the methods used to investigate exploratory search. The scope of our study is on exploratory search using metadata to a given relevant document. This describes a realistic search scenario as a majority of users visit DLs starting with a seed document which was retrieved via search engines and continue their search based on different document features of the seed document. Furthermore, our contribution in this paper is a comprehensive analysis of exploratory search in a real life digital library using mixed-methods that involve gaze data, qualitative feedback and pattern analysis to investigate the users’ behaviour during our study. Additionaly we collected transaction logs during the study that could be used for further research. Using a mixed-method approach to investigate exploratory search activities provides us with information not only on the general usage of certain search activities but also with information on the perceived relevance of certain search activities as well as a structural representation of a shared search pattern. We recruit a heterogeneous pool of users from the social sciences with different academic degrees to participate in our study. Our focus is especially on the level of search activities where we study the usage of stratagems in comparison with other search activities like querying and using document recommendations.
To the best of our knowledge, such an exhaustive study on exploratory search focusing on stratagems in digital libraries and using a mixed-method approach has not been presented yet. 
\newline

This study addresses the following three research questions:
\label{RQINTRO}
\begin{itemize}
\item{RQ 1:} What are the most frequently applied stratagems in exploratory search in a state-of-the-art DL? 
How is the usage of stratagems in comparison to other search activities like the usage of recommendations and formulating queries?
\item{RQ 2:} Is there a common search pattern shared by the majority of the participants? 
\item{RQ 3:} Are there any differences in the usage of stratagems between students and postdoctoral researchers? 
\end{itemize}

We address the first research question by examining the user journeys collected during study, the screen casts and gaze data provided by the eye tracking device. RQ 2 is addressed by using a tree graph representation of the participants search activities.
To investigate potential differences in the search behaviour of our two different types of participants (students and postdoctoral researchers) (RQ3) we perform a \linebreak non-parametric Mann-Whitney test.

The paper is structured as follows. In the next section an overview on related work is provided. In Section \ref{methodandsetup} we describe the set-up of the user study. A description of the methods used in the user study is presented in Section \ref{methodology}. Demographics of the participants are reported in Section \ref{demographics}. The results of our user study are presented in Section \ref{results} and discussed in Section \ref{discussionchap}.

%% file: relatedwork.tex
\section{Related Work}\label{relatedwork}

In the following section we discuss related work with respect to exploratory search, eye tracking studies and pattern analysis.

\subsection{Background}
Exploratory search tasks usually comprise search activities on the level of learning and investigating that go beyond simple look up tasks such as known item search \cite{marchionini2006exploratory}. Due to the complexity of exploratory search tasks they involve various search activities on the level of moves, tactics and stratagems. Xie \cite{Xie2002} and Joo and Xie \cite{Joo/Xie:13} e.g. investigated the relationship between users' search tactic selections and search outputs while conducting exploratory searches in DLs. To date numerous studies have been conducted that aim to understand users search behaviour and search activities during exploratory search. Ellis \cite{ellis1989behavioural} studied the search behaviour of social scientists and identified six generic features: \em Starting \em  (e.g. to identify a paper to start with), \em Chaining \em(e.g. to follow references in a certain paper), \em Browsing \em(e.g. to browse all papers by a certain author), \em Differentiating \em(e.g. to judge a source based on their nature), \em Monitoring  \em(e.g. to subscribe to an altering service) and \em Extracting  \em(e.g. to identify material in a well known journal). 
In Meho and Tibbo \cite{meho2003modeling}, Ellis' information-seeking behaviour model is revised by conducting structured and semi-structured e-mail interviews. Their study confirms and extends Ellis' model by four new features: \em Accessing\em, \em Networking\em, \em Verifying  \em and \em Information Managing\em. 

In our paper we are primarily interested in search activities that connect different relevant material. Thus, our main focus is on early stages of exploratory search like: Starting, Chaining and Browsing. For the present work we ignore features like monitoring or extracting as this involves a longer observation of participants and a different set-up of our study.

\section*{Studies on exploratory search}
 
In Athukorala et al. \cite{athukorala2013information} a mixed-method study is presented involving interviews, diary logs, user observations and a web survey. They recruited six participants with different academic degrees: PhD, post-doctoral and senior researchers. To validate their results, the authors conducted an online survey with a larger population of 76 computer scientists. The study consisted of three parts: a) an interview amongst others about search methods and search strategies, b) an observation phase in which the participants were observed during information seeking for a real purpose work task and c) a longitudinal diary study in which the participants were asked to keep record on their search behaviour during information search. Athukorala et al. showed that keeping up to date is the most frequent purpose of searching for computer scientists. The participants considered the exploration of an unfamiliar topic as the most challenging search task. Furthermore, they showed that backward-chaining is the most frequently used literature review technique. 

In Wildemuth \cite{wildemuth2004effects} the search tactics of medical students during a search task in a factual database are examined. The students searched a medical database at three occasions over a period of nine months. The individual moves were analysed by examining maximal repeating pattern. The results show that the most common search tactic was the specification of a concept followed by extending one or more concepts and narrowing the retrieved result set. Furthermore, they showed that domain knowledge affects the search behaviour. With more knowledge of the domain the participants changed their search tactics.

To distinguish between exploratory and lookup search tasks Athukorala et al. \cite{athukorala2015exploratory} conducted an user study with 32 participants from the computer sciences domain. The main objective of the study was to collect information on search behaviours to investigate how well different task types can be distinguished. To characterise a search session, different features were used like for instance task completion time, scroll-depth and query length. They showed that exploratory search took longer to complete, had a higher scroll-depth and involved shorter queries than in look up tasks. An analysis of gaze data showed only minor differences between look up and exploratory search tasks. 

In \cite{wilson2008longitudinal} a longitudinal study on exploratory search in a Newsfilm Online Archive called mSpace is presented. For a period of one month 22 participants took part in the study (11 known and 11 unknown online participants which were logged using the system). The aim of the study was to investigate the real-life use of exploratory and keyword styles of search. The study showed that the usage of exploratory search and keyword search was balanced throughout the study. They furthermore showed that exploratory features were \linebreak utilised to produce more expressive keyword searches. 

In Kules et al. \cite{kules2009exploratory} a study is conducted that examines how people use facets in an online public access catalogue by analysing gaze data. The main goal of the study was to learn what parts of the faceted interfaces searchers attend to, for how long, and in what order. For this reason the authors defined three areas of interest: search result pages, facets and queries. They showed that the most time is spent on inspecting the search result pages (50 sec. per task) followed by facets (25 sec.) and queries (6 sec.). An experiment comparing the search behaviour of experts and novices when searching a traditional search engine and a social tagging system is presented in \cite{kang2010exploratory}. They recruited 48 participants who were asked to solve an exploratory search task by using Google and the social bookmarking service delicious. They showed that experts used queries more often than novices while novices used more tag-based queries. The authors assume that experts are more likely to conduct queries from existing knowledge while novices rely on existing information in the environment. A qualitative study observing the usage of a DL was presented in \cite{fields2005designing}. The study starts with the observation that experienced users of DLs are more effective than non-experts. The purpose of the study was then to investigate the nature of experienced DL users in more detail in order to design interfaces that support unexperienced users.

In our user study (see Section \ref{methodandsetup}) we follow the setting outlined in previous exploratory search studies (see sections above) which compared unexperienced with expert users in a controlled set-up \cite{athukorala2013information,kules2009exploratory,kang2010exploratory,fields2005designing}. The approach taken in our study can be compared best to \cite{athukorala2013information} because both studies used a mixed-methods design. Our approach has the strength that it includes a technique to visualise user behaviour which is capable to report abstract behaviour which is not restricted to different predefined user groups.

\section*{Eye tracking studies and search behaviour}
Kelly \cite{kelly2009methods} gives an overview on potentials and limitations of eye tracking evaluations in interactive information retrieval (IIR): On the one hand, people need to sit still during the session, hence, eye tracking fits better for short search tasks. Also, these experiments create a huge amount of data that needs to be sorted out before running an analysis. On the other hand, eye tracking opens up chances for a better understanding of the user's behaviour: Gaze data provides us with detailed information about what metadata users look at and how their attention is directed during the search process.  

Although there are various studies on search behaviour on the Web that feature eye tracking such as \cite{granka2004eye, goldberg2002eye, cutrell2007you, lorigo2008eye}, this method is not yet common in digital library research, but is becoming more and more popular. 
Bierig et al. \cite{bierig2009user} give an example for a framework that incorporates eye tracking with log analyses in IIR-evaluations. Likewise Tran and Fuhr \citep{Tran/Fuhr:12b, Tran/Fuhr:12a} developed a new framework for dynamic areas of interest (aoi) as an improvement for eye tracking experiments. Buscher et al. \cite{buscher2012attentive} derived relevance feedback from eye movements for improving the quality of search result lists. Their results showed that users' reading behaviour changes with the relevance of a text or document: Though the fixation duration does not differ when the relevance of a document increases, the number of saccades decreases. Loizides et al. \cite{loizides2014interactive} investigated the reading behaviour during the search process in DLs and collected information on how users read information about documents and how their attention can be guided, resulting in design implications for custom interfaces in DLs. 


\section*{Search pattern analysis}
An important aspect for understanding user behaviour is the underlying information intention. In Mitsui et al. \cite{Mitsui:2016}, a study has been conducted in which the search intention given by the participants could be identified automatically. The authors found evidence that there is a connection between search pattern and task type. Cole et al. \cite{cole2015user} were able to distinguish between low-level tasks, based on the activity pattern and introduce a novel technique that allows to detect aspects of tasks. Going beyond the connection between pattern and task, Busher et al. \cite{buscher2012largescale} found, that the user's task influences the result page examination behaviour. They analysed queries, clicks, mouse cursor movement, scrolling, and text highlighting that was collected from the usage of the Bing search engine during a time period of 13 days. By using a set of features derived from the logged data, they were able to cluster the data into six clusters. By closer examining the task type, connections between task type and behavioural aspects were identified. By clustering only data from non-navigational tasks, they were able to distinguish three types of search engine result pages (SERP) examiners: economic, exhaustive-active and exhaustive-passive user. 
While economic users do not spend much time on SERPs, show more mouse movement, and abandon SERPs more often, users from the exhaustive groups investigate their SERPs more intensely. Similar groups have been found in \cite{aula2005eye}. The authors have conducted a lab study with 28 participants. Based on the eye tracking data, specific examination patterns were identified and manually clustered into the two groups economic and exhaustive evaluation styles. For both groups significant differences in the search behaviour could be found. White and Drucker \cite{white2007investigating} also focussed on patterns in the search behaviour. They collected five months of live data from 3290 users and extracted the users' search trails. Based on these trails, they identified differences in the interaction patterns, which led to two identifiable user groups, navigators and explorers. Navigators showed more consistent interaction patterns. They show few deviations in their behaviour, tackle problems sequentially and revisit former pages more often. In contrast, explorers use a variety of different patterns, they branch frequently, submit more queries and visit new websites more often. 
\newline

It remains to investigate in how far the groups found in \cite{aula2005eye}, \cite{buscher2012largescale} and \cite{white2007investigating} are comparable. It seems that there are users who tend to solve their information need by an economic exploration, looking at SERPs less intensely and conducting new searches more often and, that there are exhaustive users who inspect their SERPs more intense. These exhaustive users repeat similar patterns more often and navigate to previously visited information. 
Our study compares to previous studies \cite{aula2005eye,buscher2012largescale,white2007investigating} due to the focus on the fundamental behaviour of users on the web; namely an effective or an exhaustive way of accomplishing search tasks. The focus on the analysis of stratagem usage in a DL in our study (see Section \ref{methodandsetup}) and the structural representations that shows the connections of these is our main contribution.  



%% file: userStudy.tex
\section{User Study}\label{methodandsetup}
\subsection{Setup}
Our user study took place in our user lab in single sessions with a duration of about 30 minutes each. We made sure that the conditions were the same in every session. The experiment was run on a laptop connected to an external 22''-monitor as the stimulus monitor for the participants. An additional keyboard and a mouse were attached to the laptop as controlling devices. The display of the laptop was used for observation. We used an SMI iView Remote Eye tracking Device 250 that was installed at the bottom of the stimulus monitor. The screen activities as well as the eye movements were recorded by the corresponding software SMI Experiment Suite 360\degree. We used a nine point calibration with a both visual and quantitative validation to ensure the quality of the gaze data and defined a sampling frequency of 250Hz for recording the eye movements. All participants used Mozilla Firefox for working on the task.

\subsection{Scenario}
\begin{figure}
	\begin{center}
		\begin{center}
			\includegraphics[trim=60 510 285 70,clip, width=1.0\linewidth]{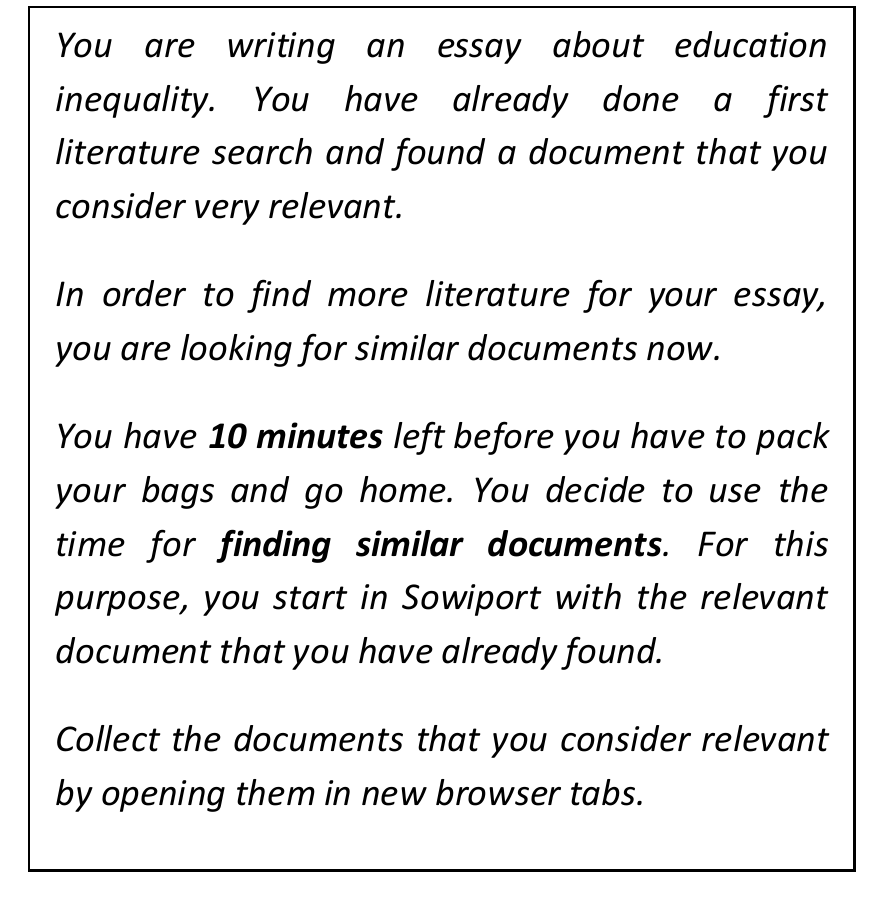}
		\end{center}
		\caption{The scenario for the user study.}
		\label{fig:scenario}
	\end{center}
	\vspace{0.0em}
\end{figure}

\begin{figure*}[htbp]
	\subfloat[Seed document] {\includegraphics[width=1\linewidth]{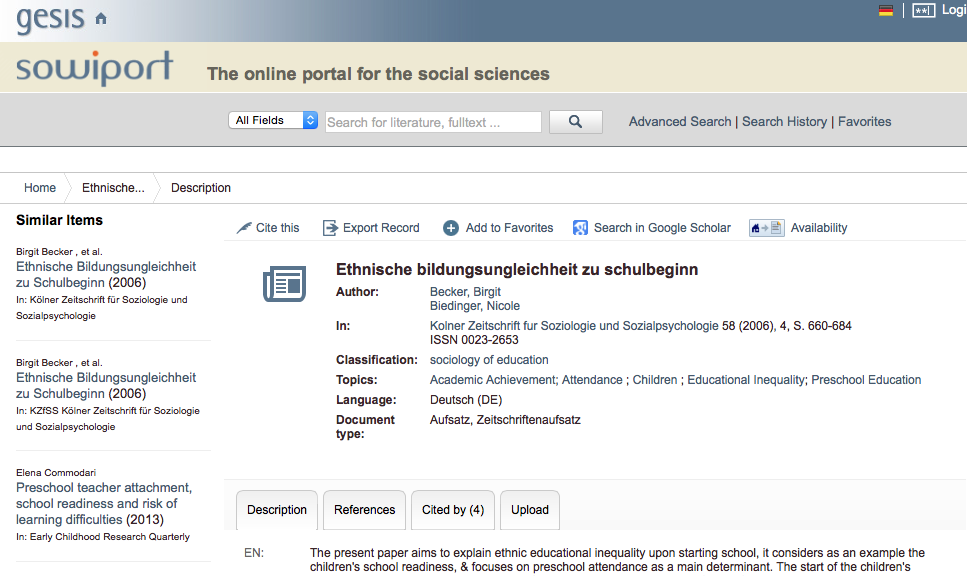}}
	\vspace{\fill}
	\subfloat[Citation tab] {\includegraphics[width=0.45\linewidth]{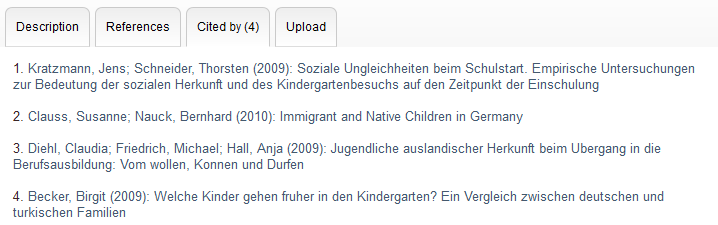}}
	\hspace{\fill}
	\subfloat[Reference tab]{\includegraphics[width=0.45\linewidth]{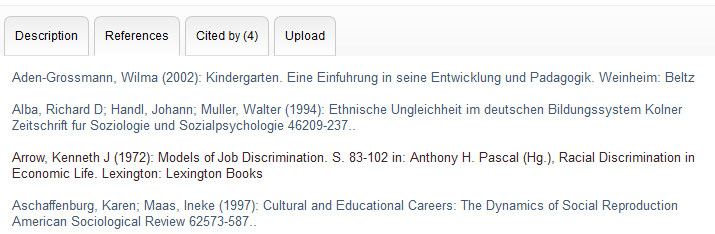}}
	\hspace{\fill}
	\caption{Seed document for the user study.}
			\label{fig:seeddocdesc}
\end{figure*}

All participants had to accomplish the same task shown in Figure \ref{fig:scenario}.
They started with the same seed document titled \em Ethnical education inequality at start of school \em (German original title: "Ethnische Bildungsungleichheit zu Schulbeginn") and had a limit of ten minutes to solve the task. Having to use Sowiport \cite{Hienert2015} for their literature search, the participants had access to about 9 million documents from 18 different databases of which six are English-language ones. 
The seed document (see Figure \ref{fig:seeddocdesc}) was published by two authors, contained five keywords and one classification term and was published in a German journal for sociology and social psychology. Each of these information were implemented as hyperlinks and could be utilised for further exploration. In separate tab views the participants could browse through citations (four citations all accessible in Sowiport via hyperlinks), reference information (70 references of which 33 are accessible in Sowiport via hyperlinks) and read the abstract of the given document. Additionally, the participants had access to the full text of the seed document. 
Besides this information the participants were provided with ten document recommendations. Five of them were provided by the SOLR \em more like this function \em while the remaining five were associated documents that were published in the same journal. As Sowiport comprises 18 different databases it is possible that duplicate documents are recommended. The seed document was chosen as it comprises a reasonable number of metadata that facilitate a further exploration and because the topic of the document would neither be too specific nor too generic for participants from the social sciences.

The aim of the study was to investigate the participants' search behaviour in an exploratory search task with a special interest in search activities on the stratagem level. We assume that these are the most frequent during early stages of exploratory search. Therefore, we limit our observation on the first three features given by Ellis \cite{ellis1989behavioural}: Starting, Chaining and Browsing. Although our time constraints are rather low (ten minutes) we assume that most search activities on these features occur during the early stages of exploratory search. Due to the complexity of exploratory search, the time to complete such a task is usually much longer than ten minutes. Therefore, the results of our search scenario only reflect certain initial aspects of exploratory search.

\subsection{Procedure}
The user study was divided into three parts. Each session followed a detailed protocol to make sure that the conditions were the same for each participant. After a short introduction to the background of the study and the procedure, each participant signed a consent form about the recording of the gaze data as well as the screen activities. 

In the first part of the study, the participants were handed out a pre-questionnaire consisting of seven questions that covered demographic information as well as their search experience in DLs. Regarding their search experience we asked for the participant's overall use of DLs and the use of Web of Science and Sowiport in particular as well as Google Scholar. We used a five-point scale (1=very rarely; 5=very often). 

In the second and main part of the user study the participants were provided with the scenario and the task they had to solve. For this, the eye tracking technology as well as the task were shortly explained. The participants were asked to take an upright and comfortable sitting position in order to move as little as possible during the search task. Then, the monitor with the eye tracker was aligned to their height and eye level so that their eyes were positioned in the center of the area the eye tracker captures. The eye tracker was calibrated and only if the result was good or very good, the experiment was started. The participants were first displayed the scenario. After reading, they proceeded to the seed document for the search task in Sowiport by pressing the space bar. While the participants were solving the task, we observed their activities and eye movements on the observation screen and noted down which stratagems they used and which information they looked at. We focused on the six stratagems that were derived from \cite{Bates:79} and that we had already used in the online survey:
\begin{itemize}
\item Follow references in the current document. 
\item Inspect the list of documents that cite the current document.
\item Search keywords that describe the current document.
\item Look for papers the authors of the current document has/have published.
\item Browse the conference/journal the current document was published in.
\item Browse a thesaurus to find  classification terms related to the current document.
\item Use document recommendations.
\item Formulate queries.
\end{itemize}
Besides stratagems we measured the usage of recommended documents and the formulation of queries. 
After ten minutes, the participants were told to end their search and review the documents they had collected in browser tabs. They were instructed to close the documents they did not consider relevant anymore. 

In the third and last part of the session, the subjects had to fill out a questionnaire covering five questions about the task in general that mostly had to be answered on five-point scales: How would you consider the overall difficulty of the given task (1=very easy; 5=very difficult)? How difficult was it to find relevant documents (1=very easy; 5=very difficult)? How would you rate the system's ability in providing you with relevant documents (1=not good; 5=very good)? How successful was your search (1=unsuccessful; 5=successful)? Which part of solving the task would you consider particularly difficult and why? Furthermore, they had to answer questions about their use of stratagems that were based on the observations we made during the search task. On the one hand we asked the participants to state the reasons for their use of a stratagem, rate its usefulness on a five-point scale and justify their usefulness rating. To avoid confusing the participants or creating the impression that the use of stratagems was expected from them and thus, influencing their searching behaviour, neither the introduction and the task nor the questionnaires mentioned stratagems. In the post-questionnaire the stratagems were paraphrased via their characteristics:
\begin{itemize}
\item You solved the task by following the references of the document.
\begin{itemize}
\item Why did you choose this method?
\item How useful was this method (1=not at all useful, 5=very useful)?
\item Please give reasons for your choice of usefulness.
\end{itemize}
\end{itemize}
On the other hand we asked why they had looked at a particular information but decided not to use it as a search method:
\begin{itemize}
\item You looked at the classification terms of a document but did not follow them.
\begin{itemize}
\item Why didn’t you use this method?
\end{itemize}
\end{itemize}
The participants were encouraged to answer the questions intuitively. After filling out the questionnaire, the session was ended and the participants were thanked.

%% file: methodology.tex
\section{Methodology}\label{methodology}
Using a mixed-method to investigate exploratory search activities provides us with information not only on the general usage of certain search activities but also with information on the perceived relevance of certain search activities as well as a structural representation of a shared search pattern. 

We examine the results of our user study using screen casts, eye tracking data and pattern analysis. Each method addresses one of the research questions described in Section \ref{RQINTRO}.
\subsection{Eye tracking}\label{eyetracking}
For each participant, full screen records were taken during the search task that showed their gaze data and also captured the navigation bar of the web browser as well as overlaying dynamic elements. We used these records to reproduce the user journey of each participant. The used search activities\footnote{If a participant combined two search activities like for example keywords and queries we only report on the first search activity.} (stratagems, queries and recommendations), the collected relevant documents and the deleted documents per stratagem were counted. 

To analyse the participants' gaze data, we needed to do some preprocessing. First, we defined a fixation time threshold of 104ms, following \cite{reichle2012using}, that marks the start of lexical processing. Second, for each participant, we determined the stable eye by comparing the scanpaths of the left and the right eye of a participant in order to utilise divergences. Third, we reduced the data to three stimuli -- the seed document, the reference list of the seed document and the citation list of the seed document -- and the first visit on each of these stimuli. We focused on these variants of the seed document, since it is the only document that each participant visited before starting their explorative and individual search through the digital library. By opening the reference list and the citation list of the seed document, the view of the document changes for the user and thus, a new variant of it is created and saved as a stimulus.
Since these three variants are the ones that cover the stratagems that are subject of this study, we concentrated on the three of them. In doing so, we focused on the timespan between entering the document –- respectively the document variant –- and the first interaction with a clickable object. Thus, overlaying elements such as the interactive search term recommender that would have distorted the gaze data were eliminated. All three stimuli also cover the first feature of Ellis's model, the Starting: During the Starting, the user gets to know the paper that forms the starting point of the search session, including also the examination of the document's references or citations. The other two of Ellis's features that this work focuses on (Chaining and Browsing) are entered after the user has utilised at least one stratagem (e.g. clicked on a reference or producing a list of search results). 
To analyse the eye tracking data we exploited heat maps in order to obtain a visual representation of the participants' gaze data and areas of interest (aoi). We quantify and compare potential search activities. The results of the usage frequencies and the eye tracking experiment are presented in Section \ref{descriptive}.
To further analyse the participants search behaviour we transformed the search activities of each participant into a tree graph representation. In \ref{patternMethod} we describe how we create this representation in detail. Given the tree graphs we address the second research question in \ref{patternAnalysis}.

In Section \ref{divinparticipants} we compare the results of the usage frequency and the pattern analysis between the two groups of participants. To seek for significant differences in the search behaviour we utilise a non-parametric Mann-Whitney test ($\alpha$ $\le$ 0.05).

\subsection{Pattern analysis}\label{patternMethod} 
In order to address our second research question we need a method to analyse our participants' sessions on a structural basis. Therefore, we derive trees as a structural representation from the sequences of our participants' search sessions. We discard the type of the activities and the order in which they are executed. In this way, we obtain an structural representation of the search sessions. We do this, to investigate how far participants explore from the seed document and how detailed they investigate different directions. This approach could be compared to analysing the patterns in which one would explore an unknown city. Although we loose much information about the search process by discarding time and type of action, this approach allows us to combine different sessions and thus to create one session tree for multiple sessions. This allows us to investigate structural characteristics of entire groups.


The user sessions within our study can be represented as a sequence of activities and visited objects in the retrieval system. Each activity (e.g. clicking on a document or triggering a query) is interpreted as a transition between two objects (e.g. a document, a citation list, a SERP). It can be denoted as $object_{a} \rightarrow object_{b}$. In our representation, the type of activity is ignored. As each session is temporally consistent, each session can be encoded in this way.

\begin{figure}[htbp]
	\subfloat[session 1]{\includegraphics[width=0.31\linewidth]{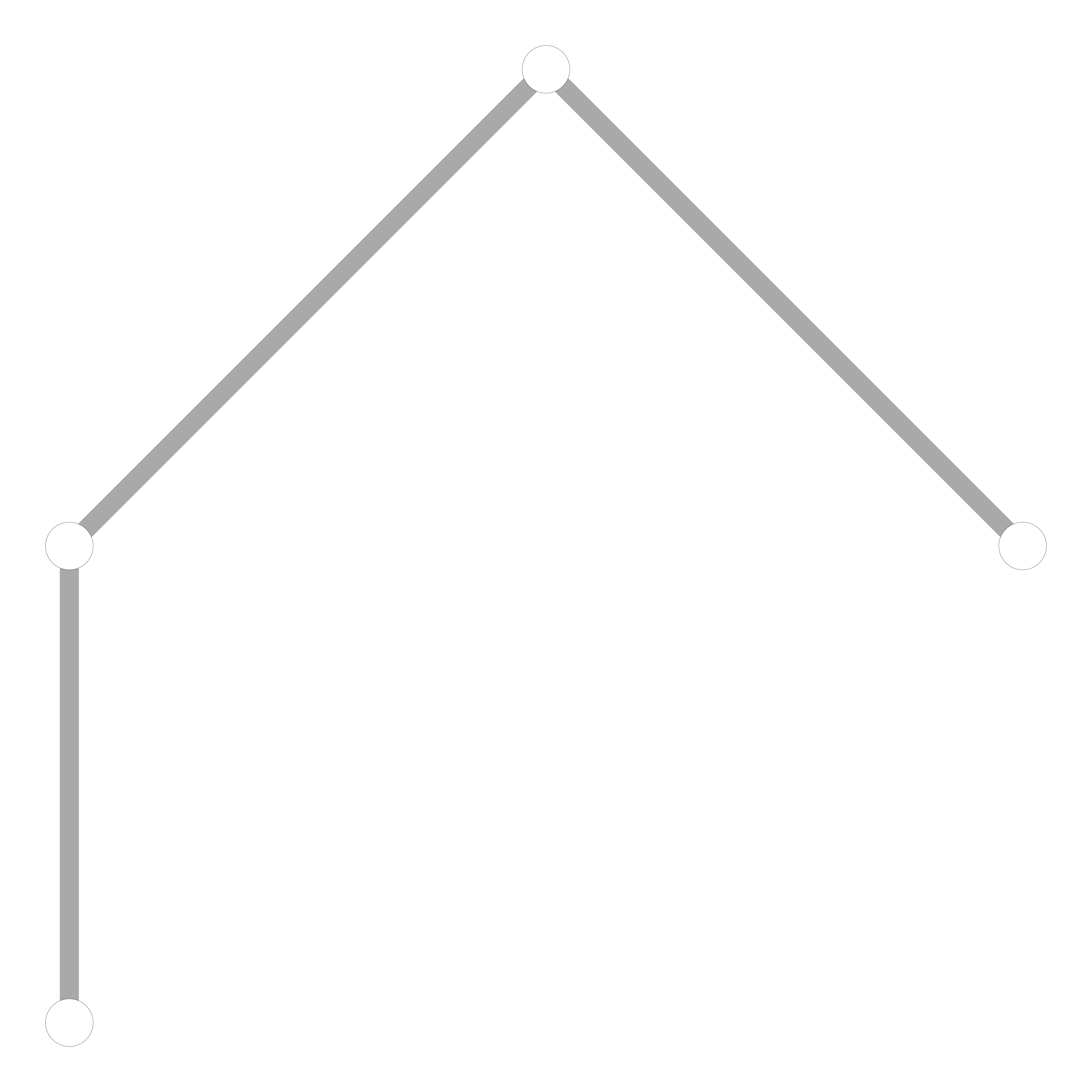}}
	\subfloat[session 2]{\includegraphics[width=0.31\linewidth]{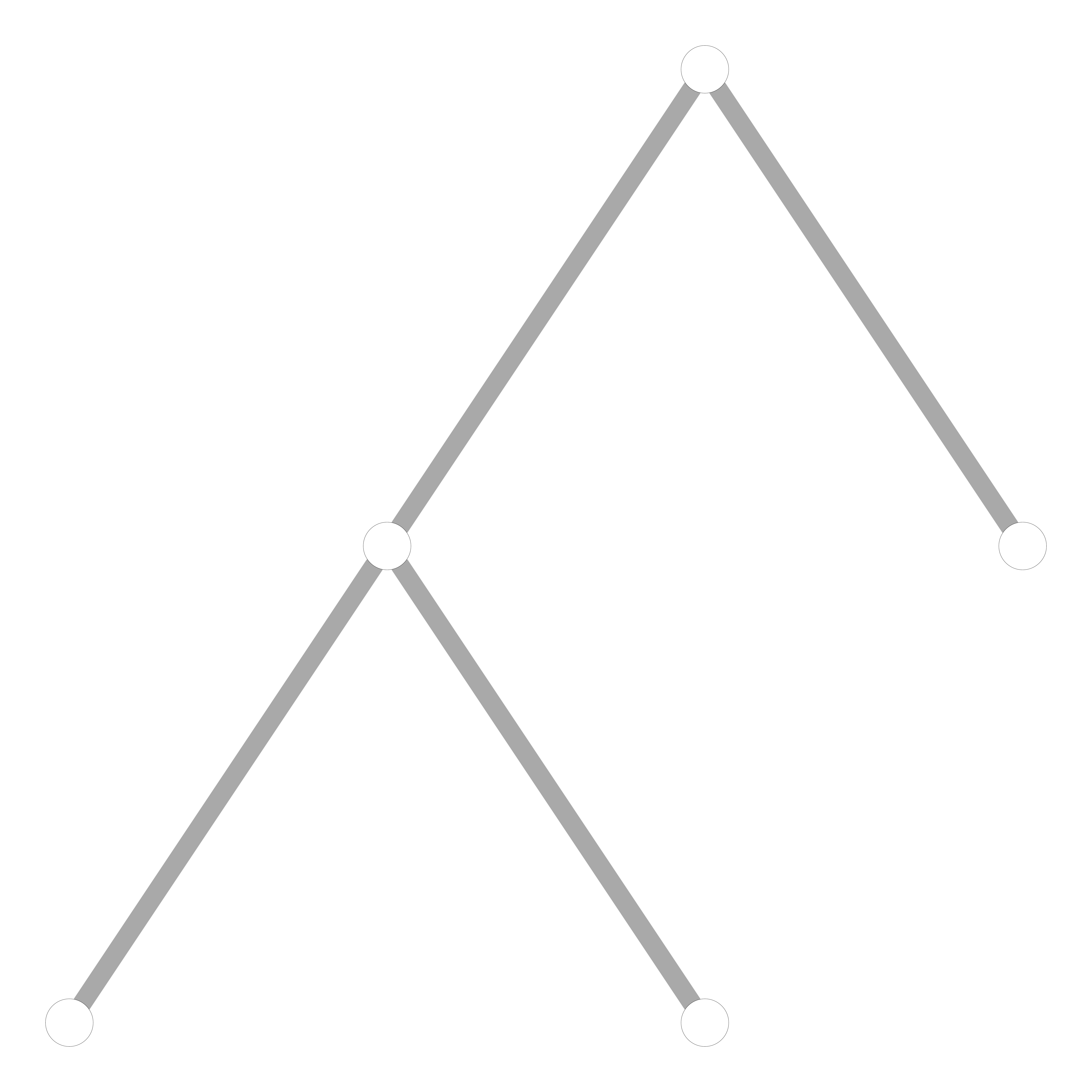}}
	\subfloat[session 3]{\includegraphics[width=0.31\linewidth]{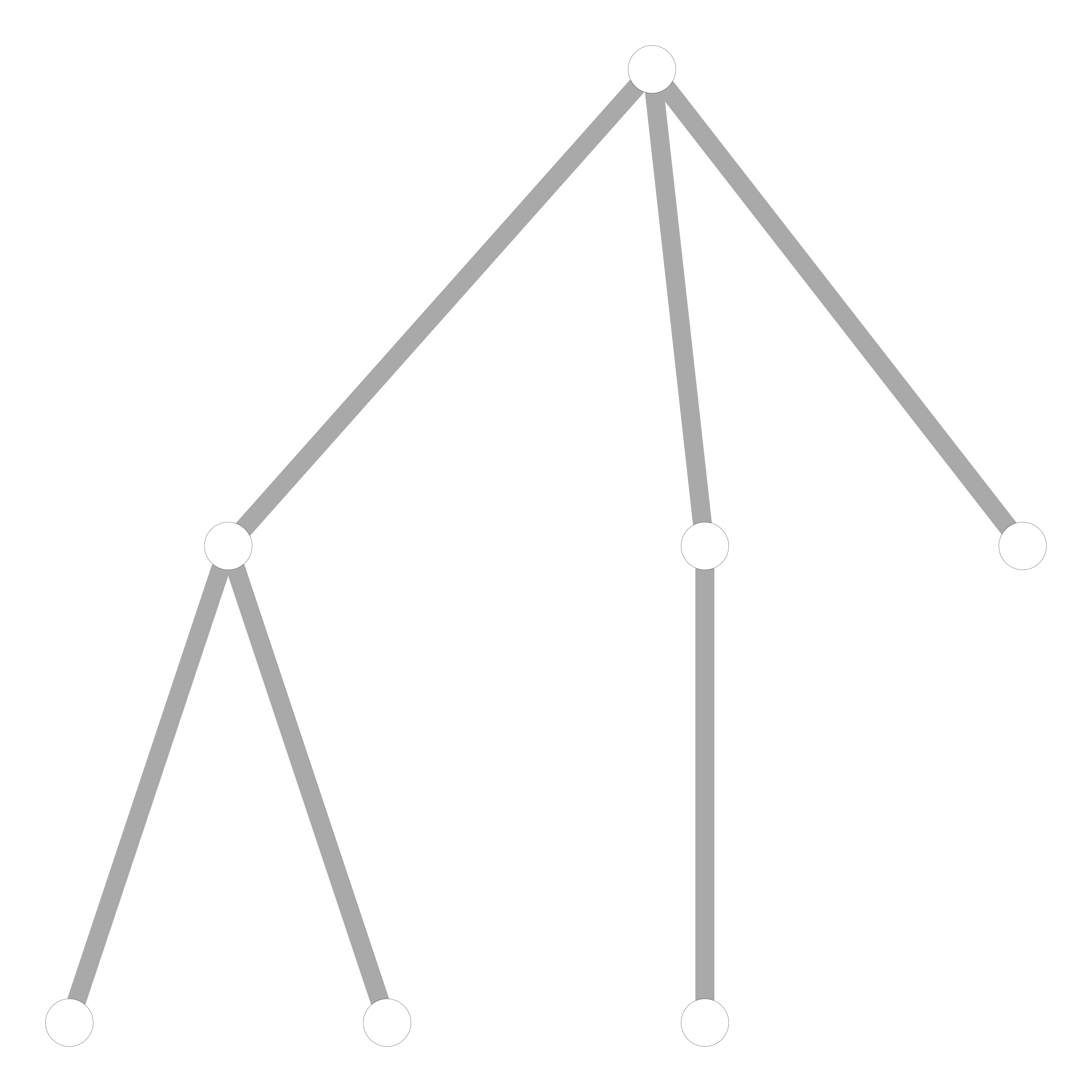}} \\
	\begin{itemize}[align=left,
		leftmargin=2em,
		itemindent=0pt,
		labelsep=0pt,
		labelwidth=2em]
		\item[\textbf{Session 1: }]  $doc_{seed} \rightarrow citation \rightarrow doc_1 \rightarrow citation  \rightarrow doc_{seed} \rightarrow search$	\\	
		\item[\textbf{Session 2: }] $doc_{seed} \rightarrow journal \rightarrow doc_{seed} \rightarrow author \rightarrow doc_1 \rightarrow author \rightarrow doc_2$	\\
		\item[\textbf{Session 3: }]  $doc_{seed} \rightarrow search \rightarrow doc_1 \rightarrow search \rightarrow doc_2 \rightarrow search \rightarrow doc_{seed} \rightarrow journal \rightarrow doc_{seed} \rightarrow citation \rightarrow doc_3$	
	\end{itemize}
	\caption{Session trees (a-c) for three example search sessions (1-3)}
	\label{fig:tree_examples}
\end{figure}

To illustrate how we create such session trees, Figure \ref{fig:tree_examples} shows three fictitious examples of session trees and the corresponding search patterns. In the first session starting from the seed document, the document's citations were clicked and one of the cited documents visited. This is represented by the left two nodes in session tree \ref{fig:tree_examples}a. Each action (e.g. click on citation list, search) is represented as an edge and each object (e.g. citation list, SERP) as a node. As each action results in an object, we always insert edge and node together. Activity and result objects are closely coupled. After that, the user returns to the citation list and then to the seed document. As this only involves already visited pages, it does not result in additional nodes. At the end of the session, the user conducts a search which accounts for the right node in session tree a.

For this representation we ignore the type of the activities and objects. The information whether a node represents a SERP, or a citation list, or an edge represents a search or a click on citations, is discarded.  Also, we treat the usage of recommendations equal to other stratagems. If at least one recommendation is clicked, we insert a node representing the list of recommendations and a child node for each clicked recommendation, although no extra click is needed to display the list of recommendations. Furthermore, we discard the order of the search session. Instead, we sort the tree by subtree size from left to right. Therefore, on one level, nodes with more subnodes are sorted further to the left than those with less subnodes. This can be observed in session 2 of Figure \ref{fig:tree_examples}. The usage of the author stratagem appears after the journal usage, but the corresponding nodes are on the left side, because there are more subsequent actions involved.

Session trees allow a structural overview of the way in which a participant has searched. One can see whether the participant has inspected a few lists of documents intensely (e.g. SERP or citation list) by clicking on many documents or has preferred to browse less through a higher amount of different lists, while only investigating a smaller share of documents. 

\begin{figure}[htbp]
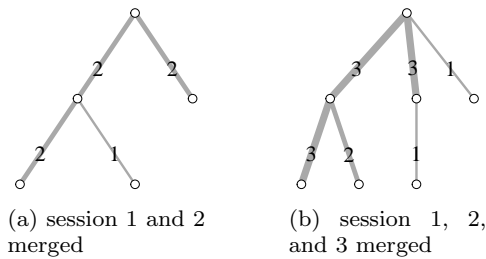

	\hspace{\fill}
	\subfloat[session 1 and 2 merged] {\includegraphics[width=0.31\linewidth]{example_merged_1-2.pdf}}
	\hspace{\fill}
	\subfloat[session 1, 2, and 3 merged]{\includegraphics[width=0.31\linewidth]{example_merged_1-2-3.pdf}}
	\hspace{\fill}
	\caption{Combined session trees of the example sessions from Figure \ref{fig:tree_examples}.}
	\label{fig:tree_examples_merged}
\end{figure}

To extend the analysis from individuals to groups of participants and thus, to understand and find common patterns, we need a way to combine multiple session trees into one conjoint tree. To do this, we have implemented a tool to combine the session trees.\footnote{https://github.com/wilkovanhoek/amur-session-tree/ijdl} Instead of combining all trees at once, we merge pairs of trees iteratively. We start with an empty tree and merge it with the first session tree. The resulting tree is then merged with the next tree and so forth. Figure \ref{fig:tree_examples_merged}a shows the merged tree of the example session tree 1 (cf. Figure \ref{fig:tree_examples}a) and session tree 2 (cf. Figure \ref{fig:tree_examples}b). When two edges and their nodes are merged, the edge weights are summed. Initially each edge has a weight of 1. The higher the edge weight is, the more often it has been merged. Thus, the higher an edge weight is, the more session trees it represents. The edge-thickness in the graph represents the edge weight. After merging session tree 1 and 2, we merge the resulting tree with the tree of session 3 (cf. Figure \ref{fig:tree_examples}c), shown in Figure \ref{fig:tree_examples_merged}b.

When merging two trees, one has to decide which edges and nodes have to be merged. Each node and the node's child nodes can be treated as an individual subtrees. The decision whether to merge two edges is therefore transferred to the merging of their nodes' subtrees and so on. For each pair of edges on one level, we create all possible combinations of subtrees and select the best combination along two conditions. First, the number of nodes in the resulting subtree has to be minimal. Second, the weight distribution of the subtree has to be optimized. 

For the second condition, we need to define what a weight of a subtree is and what an optimum for the distribution of the subtree weight is. We define the subtree weight $W(p)$ as follows: Let $p \in N_T$ be a node in the tree $T$, $w(p)$ be the weight of the edge leading to the node $p$, and $C_p$ the set of child nodes of $p$. The weight of the subtree with the root node $p$ is then defined as:

\begin{equation}
W(p) =
\begin{cases}
	\lg(2 \cdot w(p)) & \text{if } C_p=\emptyset\\
	\lg( \sum_{q \in C_p} w(q) \cdot W(q)) & \text{else } \\
\end{cases}
\end{equation}

First, we sum the product of the weight of each edge leading to a child node of $p$ and the subtree weight of that child node and than, calculate the logarithm of this sum. We use the logarithm only to keep the subtree weight from increasing exponentially, as we only need it to compare the subtrees, not to assess an actual summed subtree weight.

After defining the subtree weight, we need to define what an optimal subtree weighting is. Because we want the weight of the resulting subtrees to increase from left to right and we do not want the weight to be distributed equally, we favour building maximal heavy subtrees. Therefore, the merge with the heaviest subtree is considered to be optimal.

Following the described procedure, we receive exactly one resulting tree, when combining two session trees. However, when merging multiple trees iteratively, the resulting tree depends on the order in which the trees are merged. To find the optimal merging order, we would need to calculate all possible permutations in which the session trees could be merged. This would result in $32! \approx 2,63 \mathbin{\times} 10^{35}$ merging orders for the complete set of trees in our study. As this exceeded our computational capacities for this paper, we decided to use another approach. Before merging the trees, we sort the session trees in ascending order with respect to their root node's subtree weight. Now, when merging them, the resulting tree of each merge slowly increases in weight (in general trees with a smaller subtree weight tend to be more compact trees). In this way, common structures that exist in many trees are merged very early, whereas outliers are merged later. Note that the sorting does not guarantee an optimal merging order. However, it ensures that there is only one resulting tree for a set of trees and that all combined trees are created with the same procedure. We believe, that sorting the trees results in a suboptimal tree, that is close enough to the optimal tree to allow to draw meaningful conclusions for our analysis.

%% file: demographics.tex
\section{Demographics}\label{demographics}
The user study involved 32 participants that were divided into two groups A and B. While group A included 16 students, group B consisted of 16 postdoctoral researchers. All participants were social scientists from different fields of study that had been recruited via e-mail and personal recommendation.

The students were aged 22 to 35 years (m=26.38, sd=3.76). 75\% were female and 25\% male. About 19\% of the students had no academic degree yet, 69\% held a bachelor's degree, 15\% held a master's degree and 6\% had a diploma.
The postdoctoral researchers' age was slightly higher and ranged from 30 to 62 (m=40.19, sd=9.23). In this group, 50\% were female and 50\% male.

The participants were asked about their search expertise, in particular their experience in searching digital libraries and rated it on five point  scales (1=none at all, 5=expert).
Regarding their overall search experience in digital libraries, 25\% of the students stated to have little experience, 56\% rated their experience as moderate and 19\% considered their experience high, resulting in an average search experience of 2.94 (sd=0.66). In contrast, 38\% of the postdocs considered their experience little and 31\% moderate, while 25\% stated to have a high or expert experience, with an average search experience of 2.88 (sd=1.11).

In accordance with the online survey \cite{carevic2016survey}, we asked the participants about their use of digital libraries and Google Scholar on a five point scale (0=very rarely, 5=very often). In the students' group, 5\% stated that they used DLs in general often or very often (m=2.88, sd=0.99). 25\% indicated to use Sowiport often or very often (m=2.81, sd=1.07) and 62.5\% used Google Scholar often or very often (m=3.69, sd=0.92). 19\% of postdoctoral researchers replied to use DLs in general often or very often  and 6\% stated to use Sowiport often or very often, while 63\% claimed to use Google Scholar often or very often.

%% file: usagefrequency.tex
\subsection{Search activity usage and eye tracking}\label{descriptive}
In the following section we report descriptive statistics on the usage frequencies of search activities and gaze data provided by the eye tracking device.

\subsubsection{Search activity usage frequency}
The total usage frequency of the eight potential search activities is displayed in Table \ref{stratagemusagefrequency}. In total, the participants applied a stratagem search 137 times. The most frequently used stratagems were: using keywords, following references and citations. A journal run has only been utilised by two participants. The usage frequency of recommendations and queries is displayed in the last two rows of Table \ref{stratagemusagefrequency}. It can be seen that queries and recommendations (n=111) are less utilised than stratagems. Surprisingly only 22 participants used queries which means that ten participants solely used stratagems or recommendations to solve the task.
 
A possible explanation for the strong reliance on stratagems can be found in the qualitative feedback of the participants. Ten participants named little effort as a criterion for using a stratagem and preferred quick and easy steps to find more relevant results during the task. 12 participants named search habits and time restrictions as reasons for using or not using a stratagem. Some of them stated to have an order in which they use stratagems during research and told us that they would have used another stratagem, if there had been enough time. They also mentioned that inspecting the list of references and journal run would have cost them too much time, so that they decided not to use these stratagems during this particular search task. Additionally, the users named several document-related factors that constitute on the metadata of the respective document. 15 participants stated that they did not perform a journal run, because the regarding journal was too general for the task. Corresponding to that, 12 participants told us that the specific author was crucial for their decision to search for an author. If the participants surmise or know that an author focuses on the topic of interest, they are more likely to perform an author run. Regarding the classification terms and the keywords, six respectively three participants said that they only use them for further search if they are relevant for the topic and neither too general nor too specific.

\begin{table}[h]
\centering
\caption{Total usage frequency of the six stratagems in comparison to queries and recommendations.}
\label{stratagemusagefrequency}
\begin{tabular}{p{1.4cm}p{2.5cm}P{1.4cm}P{1.8cm}}
Type&Search Activity & Usage frequency & \# Participants\\\hline
Stratagem &Keywords &  50 & 16\\
&References & 27 &18\\
&Citations & 26 & 18 \\
&Author &  25 & 13\\
&Classifications & 5 & 5\\
&Journals& 4 & 2\\ 
   & \textbf{total stratagem}& \textbf{137} \\\hline
Other &Queries & 69 & 22\\
& Recommendations  &42 & 16\\
 & \textbf{total other}& \textbf{111} 
\end{tabular}
\end{table}

If a participant utilised a certain search activity during the post-questionnaire he or she was asked to rate the usefulness of that particular search activity on a five point scale\footnote{Not all participants responded to each search activity which results in minor differences between the actual usage of a search activity (see Table \ref{stratagemusagefrequency}) and the usefulness rating.}.  The results are displayed in Table \ref{usefulness}. The participants considered references, classifications and authors as the most useful search activities for the given task, while the usefulness of  journals, recommendations and queries was comparably low. The reasons for the usefulness ratings correspond to the reasons for the usage or rejection of a search strategy. The participants explained that references gave them a quick access to a large set of documents that were topically related. Critical aspects on references were that they provide often large lists and are time-consuming, and that some of the documents were too specific. Similar responses were stated for citations which also provided a quick access to related documents. A frequently stated critic regarding citations was that the number of citations was rather low (4 citations in our seed document). Most participants that followed the authors rated them as useful, because they had published similar and more recent documents on the same topic. Frequent comments on the usefulness of keywords were a precise search on topically related documents and a good starting point to the given subject. Topical mismatches and broadness were mentioned as downsides by the participants. The journal was stated to be too general for the topic at hand.
Positive aspects on the usage of queries were a feeling of control and a quick way to select topically adequate (neither too broad nor too narrow) documents while downsides were that the number of results was too high.

\begin{table}[h]  
\caption{Usefulness of search activities.}
\label{usefulness}       
\begin{tabular}{p{1cm}P{2.2cm}ccP{0.3cm}P{0.3cm}P{0.7cm}}
Type&Search activity &N & M&SD&Mdn&Mode
\\\hline
Stratagem&References &16& 4.19&0.7&4&4 \\
& Classifications &5 & 3.80 & 0.75&4&3\\
&Authors & 12& 3.75 & 0.92&4&4\\
&Citations &16& 3.69 & 0.98&4&4\\
& Keywords& 16&3.56 & 0.86&4&4\\
& Journal& 2&3& 0&3&3\\
\hline
Other& Queries&21 & 3.38 & 0.72&3&3\\
& Recommendations &16 & 3.19 & 1.01&3&4\\
\end{tabular}
\end{table}

\subsubsection{Stratagem usage frequency by session step}
Throughout the study the participants changed their search tactics several times. In Table \ref{usageperwindow} the percentaged usage frequencies for eight search activities are displayed taking only the initial ten search activities into account. In this section we refer to a \em step \em as an individual search activity performed by a participant on the stratagem, query or recommendation level.    
\begin{table}
\caption{Percentaged usage frequency of search activities within a session limit of ten steps.}       
\label{usageperwindow}
\begin{tabular}{p{1.2cm}p{2.2cm}p{0.6cm}p{0.6cm}p{0.6cm}p{0.7cm}}
Type&Search activity & Step 1& Steps 2-4 & Steps 5-7 & Steps 8-10 \\ \hline
Stratagem&Keywords & 21.8\%&16.1\% &18.7\%&33.3\%\\
&Citations & 21.8\% & 8.6\% & 10.9\% &8.3\%\\
&References & 18.7\% &11.8\% &3.1\%&13.8\%\\
&Author & 3.1\%&10.7\% &14.0\%&0\%\\
&Classifications & 3.1\%&3.2\% &1.5\%&0\%\\
&Journal & 0\%&3.2\% &0\%&0\%\\\hline
Other&Queries & 15.6\%&29.0\% &37.5\%&33.3\%\\
&Recommendations & 15.6\%&17.2\% &14.0\%&11.1\%\\
\end{tabular}
\end{table}

The two most frequent initial search activities starting from the seed document were using keywords or citations both applied by 21.8\% of the participants. Journals (none of the participants), Classifications and Authors (both 3.1\% of the participants) were the least frequently applied search activities. With increasing session steps we can observe a change in the participants search strategy. Queries which were only used by 15.6\% 
of the participants in the first session step were the most frequently applied search activity in the remaining session steps. The other search activities vary within the session steps like for instance the author run and the usage of keywords. 

In the post-questionnaire eight participants stated that the success of their previous search influenced their stratagem use. If they had not found a satisfying number of relevant results yet, they would have been more likely to use different stratagems than they had before. Another user-related factor mentioned by five participants was diversity of search results. These persons avoided especially searching for authors in order to increase the variety of their search results. Furthermore, several participants stated that during document search they followed an order of search strategies, based on their search expertise. This may also be an explanation for the utilisation of references and citations - the two most useful rated stratagems - at the beginning of the search session.

%% file: eyetracker.tex
\subsubsection{Eye tracking results}\label{eyetracker}
\begin{figure}[h]
\centering
 \includegraphics[width=0.45\textwidth]{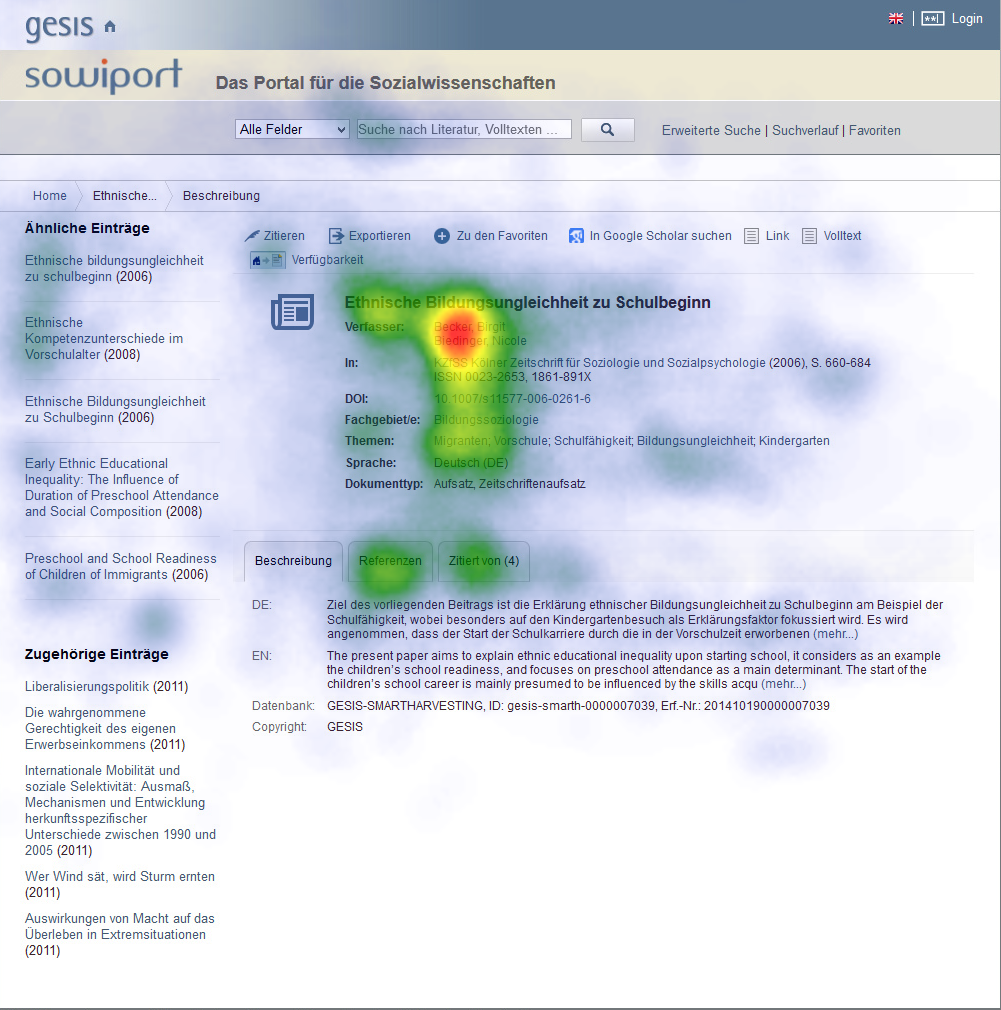}
\caption{Heat map visualization for 32 participants on the seed document.}
\label{heatmapall}
\end{figure}

The usage frequency of certain stratagems depends on the task which was to find similar documents to a given seed document. As pointed out in \cite{bates1989design} and recently in \cite{carevic2016survey}, the content of a journal may be too broad to discover something similar and therefore influence the decision on using a journal run as a search activity. In this section we investigate the perceived usefulness of the six stratagems by looking at gaze data provided by the eye tracker. This provides us with insights on the usefulness of certain stratagems without considering the underlying content. 
Figure \ref{heatmapall} shows a heat map visualization of gaze data and its distribution in the seed document for the first stimulus before any interaction with the system was performed. Giving the heatmap we can see a high frequency along the structural metadata, most intensely in the area of authors and publishing information of the seed document. Furthermore, we can recognize that the distribution is mostly focused on the first entries of the metadata, e.g. the five keywords of the seed document and decreases for the keywords. Other areas with high fixations were: classifications, citations, references, access to fulltext and recommendations. Surprisingly, the distribution along the abstract of the seed document was rather low. 

\subsubsection{Areas of interest}
\begin{figure*}
 \includegraphics[width=1\textwidth]{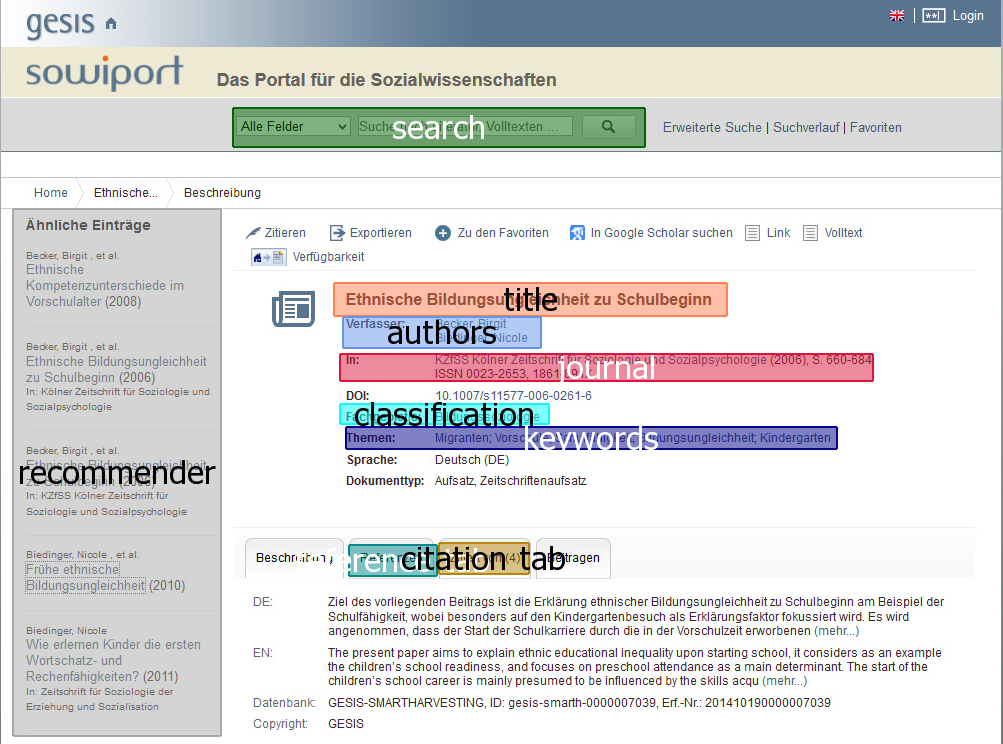}
\caption{Areas of interest for the given seed document.}
\label{aoidef}
\end{figure*}

In the following section we quantify the fixation on potential search activities by defining areas of interest (aoi) which are displayed in Figure \ref{aoidef}. Our main focus is on stratagems; therefore, we concentrate the aoi on keywords, classifications, publishing information (journal or conference proceedings), authors, citations and references. Alongside these stratagems we were interested in the fixations along query formulation and recommendations.

Given the aoi displayed in Figure \ref{aoidef} we measured the following two features: 

\begin{itemize}
\item Dwell time: The time spent on a certain aoi during the first visit.
\item Number of fixations: The total number of times a certain aoi was examined by a participant.
\end{itemize}
In order to keep the analysis of the aoi reasonable we only take the three stimuli of the seed document and the time during the entry and the first interaction on each stimulus into account. The results of the aoi are displayed in Table \ref{aoires}.

\begin{table}[h]
\caption{Number of fixations and dwell time on aoi in the seed document.}       
\label{aoires}
\begin{tabular}{p{1.2cm}p{2.2cm}p{0.3cm}P{1.3cm}P{1.6cm}}
Type & Search Activity&N&Total fixations &  Total dwell time (s) \\\hline
Stratagem&Keywords&16 & 72 (m=4.50)&22.18 (m=1.38)\\
&Classifications &18& 62 (m=3.44) & 17.82 (m=0.99)\\
&Journal &18& 98  (m=5.44)  &31.41 (m=1.74)\\
&Author &29& 165 (m=5.68) & 45.93 (m=1.58)\\
&Citations&10 & 122 (m=12.20)&33.45 (m=3.34)\\
&Reference&13& 2324 (m=178.76) & 748.74 (m=57.59)\\\hline
Other &Recommendations&14 & 176 (m=12.57) & 56.57 (m=4.0)\\
&Search&9 & 46 (m=5.11)  &10.52 (m=1.16)\\
&Title&26 & 129 (m=4.96)  &31.56 (m=1.21)\\
\end{tabular}
\end{table}

The most frequently fixated aoi were references \linebreak (2324, m=178.7), recommendations (176, m=12.5) and author information (165, m=5.6), while other areas like the search bar (46, m=5.1) or classifications terms (62, \linebreak m=3.4) were rather low. Regarding the dwell time, the participants spent the most time on inspecting the references (748.7 s, m=57.5) and the recommendations (56.5 s, m=4.0). It should be noted that citations and references are both displayed in a separate view within Sowiport. Therefore, these two aoi were measured separately. 
The numbers provided by the eye tracking device show that the usage frequency of search activities varies between the ones provided by the gaze data. The total usage frequencies of search activities by the participants in comparison to the gaze data is illustrated in Table \ref{usagevsaoi}. Although queries and recommendations were frequently utilised search activities their perceived relevance is rather low. Comparing to this the search activities on the stratagem level are more frequently studied by the participants. A noticeable difference between the actual usage and the gaze data is found for the journals and the classifications. Both were utilised only four times by the participants.

The reasons behind this can also be found in the qualitative answers from the post-questionnaire. Being asked, why they read metadata, but did not utilise it, several participants stated that metadata like authors, keywords and classification terms are general information that are important for getting to know a document, regardless of their utilisation during the search session. Regarding the journal, as stated before, the majority of the participants explained that the journal of the seed document was too general for being useful for the search.

\begin{table}[h]
\caption{Comparison of the actual usage of search activities and gaze data in the seed document.}
\label{usagevsaoi}       
\begin{tabular}{lp{2.2cm}P{0.8cm}P{0.8cm}P{1.4cm}}
Type&Search Activity&Utilised by& Fixations&Dwell Time (s)\\\hline
Stratagem&Keywords&7&72&22.18\\
&Author&1&165&45.93 \\
&Classifications&1&62&17.82\\
&Journal&0&98&31.41\\
&References&6&2324&748.74\\
&Citation&7&122&33.45\\\hline
Other&Search&5&46&10.52\\
&Recommendations&5&176&56.57\\
\end{tabular}
\end{table}


%% file: graph.tex
\subsection{Pattern analysis}\label{patternAnalysis} 

\begin{figure}[]
	\includegraphics[width=1.0\linewidth]{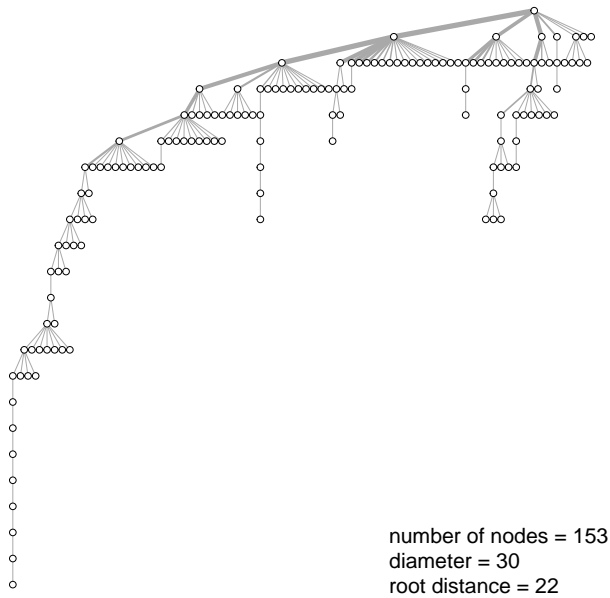}
	\caption{Combined session tree for all participants.}
	\label{fig:all_graph}       
	
\end{figure}

Figure \ref{fig:all_graph} shows the combined session tree for all participants. Taking the thickness of the edges following from the seed node into account, one can see that many user conducted at least three different actions starting from the root node. In addition, a larger group of users followed one longer trail, consisting of multiple consecutive actions (cf. trail of nodes on the left). Overall, the combined session tree is not very compact. There is a certain density within the first two levels, but below that, only a few longer trails exist in the tree. Instead, there are many edges with a small weight that are introduced by a few very intense sessions. We consider this to be noise, because it inflates the combined session tree. Therefore, we will introduce a way to reduce this noise, without removing the session tree that is responsible for the noise.

\begin{figure}[]
	\includegraphics[width=1.0\linewidth]{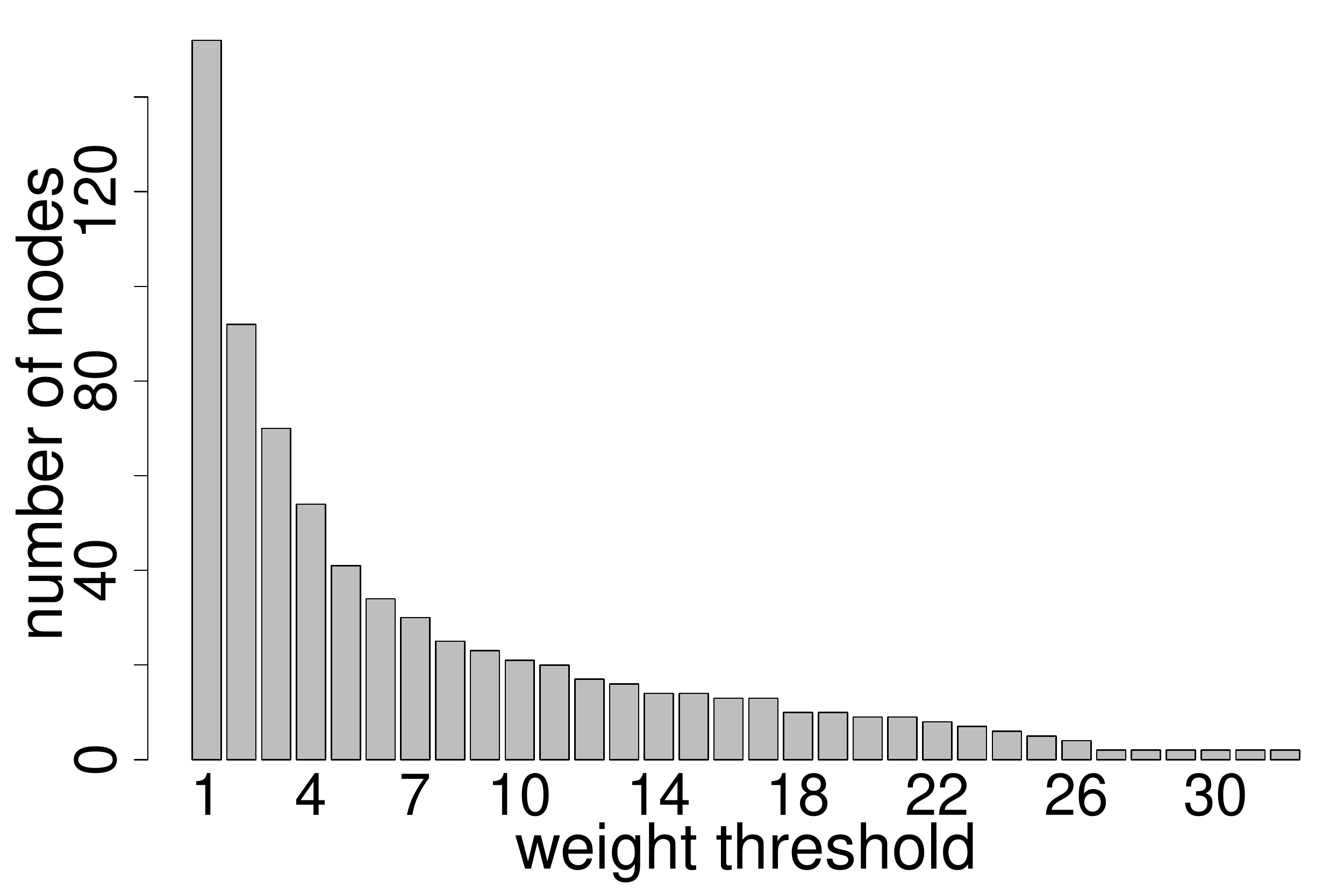}
	\caption{Number of nodes remaining in the combined session tree, when removing edges with a weight less than a given threshold.}
	\label{fig:all_graph_dist}       
\end{figure}

Edges with a low weight as displayed in Figure \ref{fig:all_graph} represent an activity that has happened in a minority of sessions. Those edges account for a high amount of nodes, although they do not give us more insight to common patterns within our data. Figure \ref{fig:all_graph_dist} illustrates how many edges of the tree have a weight above an increasing weight threshold. After the strong decline in the beginning we can observe a slower decline from a threshold of 6 and onwards. Starting at a threshold of 11 we can observe an even slower decline and some smaller plateaus, where the number of nodes stays the same for multiple thresholds and only drops a little between those plateaus. 

\begin{figure}[htbp]
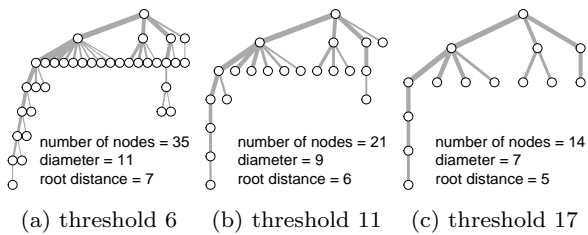

	\subfloat[threshold 6]{\includegraphics[width=0.31\linewidth]{all_graph_threshold_6.pdf}}
	\subfloat[threshold 11]{\includegraphics[width=0.31\linewidth]{all_graph_threshold_11.pdf}}
	\subfloat[threshold 17]{\includegraphics[width=0.31\linewidth]{all_graph_threshold_17.pdf}}
	\caption{Subtrees extracted from the combined session tree (cf. Figure \ref{fig:all_graph}). Each subtree is created by removing all edges with a weight below the specified threshold.} 
	\label{fig:all_threshold_trees}
\end{figure}

Figure \ref{fig:all_threshold_trees} shows three trees with different thresholds extracted from the combined session tree in Figure \ref{fig:all_graph}. Each tree is created by eliminating all edges and nodes where the edge weight is below a given threshold. We will call those trees subtrees. Figures \ref{fig:all_threshold_trees}a and \ref{fig:all_threshold_trees}b display the subtree for the thresholds 6 and 11 mentioned before. In addition, we include the subtree for a threshold of 17 which represents all nodes that appear at least in half of all session trees.

In Figure \ref{fig:all_threshold_trees}c we can see the observation of Figure \ref{fig:all_graph} very clearly. Most users start at least three independent activities from the root document and follow at least one longer trail. Comparing the three subtrees in Figure \ref{fig:all_threshold_trees}, we can observe that with an increasing threshold mainly the number of nodes per level decreases, whereas the overall structure does not change decisively.

%% file: diversity.tex
\subsection{Diversity in participants}\label{divinparticipants}
In the following section we aim to reproduce the results of our online survey \citep{carevic2016survey} which indicated a difference in search behaviour between different types of respondents. The user study was made up of two groups of participants: a) students in the social sciences and b) postdoctoral researchers in the social sciences. By comparing the results of the previous sections for the two groups we seek for significant differences in the search behaviour of the participants.
\subsubsection{Stratagem frequencies}
In Section \ref{descriptive} we report on the usage frequencies of stratagems, queries and recommendations. Utilising a non-parametric Mann-Whitney test ($\alpha$ $\le$ 0.05) we now seek for significant differences in the usage frequencies between students and postdoctoral researchers. 
The results are displayed in Table \ref{tab:multicol}. Although we found several differences between the two groups the general usage frequency of search activities does not differ significantly.

\begin{table}[h]
\caption{Comparison of students and postdoctoral researchers.}
\label{sigusagefrequency}       
\begin{center}
\begin{tabular}{|p{2.3cm}|P{1.0cm}|P{1.0cm}|P{2.5cm}|}
    \hline
& \multicolumn{3}{|c|}{Usage frequency}\\\hline
      Search activity&Students&Postdocs &Test \\\hline      
	 Keywords&  15 (N=7) & 35 (N=9) &U=106.0 p=0.37 \\   
	 Journal&  0 (N=0)& 4 (N=2) & - \\
	 Classifications& 2 (N=2) &3 (N=3) &U=120.0 p=0.63\\
	Author& 9 (N=7) &16 (N=6) &U=126.5 p=0.94\\
	Citation& 17 (N=9)& 9 (N=9) &U=110.0 p=0.45\\
	References& 17 (N=8) & 10 (N=10) &U=119.0 p=0.71\\
	Queries& 36 (N=14) & 33 (N=8) &U=100.5  p=0.29\\
	Recommendation& 24 (N=10)& 18 (N=6)&U=100.5  p=0.26\\\hline
	Total& 120 &  128&\\\hline
\end{tabular}
\end{center}
\label{tab:multicol}
\end{table}

\subsubsection{Stratagem gaze data}
In Section \ref{eyetracker} we report on gaze data for the seed document. Again, we utilise a non-parametric Mann-Whitney test ($\alpha$ $\le$ 0.05) to look for significant differences in the gaze data between students and postdoctoral researchers. In particular, we compare the dwell time and the number of fixations generated for the seed document's areas of interest (aoi). The results are displayed in Table \ref{sigusagefrequencyeytracker}. We found significant differences between the two groups in the aoi: journal. The postdoctoral researchers spent more time on the journal aoi and fixated it more often than the group of students. 
   \begin{table*} 
\caption{Comparison of gaze data between students and postdoctoral researchers.}
\label{sigusagefrequencyeytracker}
\begin{center}
\begin{tabular}{|l|c|c|c|c|c|c|c|}
    \hline
& \multicolumn{3}{|c|}{Dwell Time (s)}&   \multicolumn{3}{|c|}{Fixations}\\\hline
      AOI&Students &  Postdocs&Test& Students & Postdocs&Test\\\hline      
  Author& N=13 m=1.05 & N=16  m=2.01& u=60.00 p=0.054&  m=4.92& m=6.31&u=72.50 p=0.16\\      
  Journal& N=9 m=0.63 & N=9  m=2.85& u=12.00 p=0.012*&  m=2.55& m=8.33&u=14.5 p=0.02*\\
  Classifications& N=9 m=0.70 & N=9  m=1.27& u=24.00 p=0.14&  m=2.88& m=4&u=31.00 p=0.38\\ 
  Citations& N=5 m=3.65 & N=5  m=3.03& u=6.00 p=0.17&  m=14& m=10.4&u=7.50 p=0.28\\
 References & N=5 m=38.00 & N=8  m=69.83& u=18.00 p=0.77&  m=116.4& m=217.7&u=18.00 p=0.76\\
  Keywords& N=6 m=1.27 & N=10  m=1.45& u=28.00 p=0.82&  m=5.33& m=4&u=21.50 p=0.33\\\hline      
   Recommendations& N=9 m=2.31 & N=5  m=7.14& u=15.00 p=0.31&  m=7.7& m=21.2&u=17.50 p=0.48\\
   Search& N=6 m=0.97 & N=3  m=1.55& u=5.00 p=0.30&  m=4.83& m=5.66&u=7.50 p=0.69\\
   Title& N=14 m=0.90 & N=12  m=1.57& u=56.00 p=0.15&  m=4& m=6.08&u=59.00 p=0.19\\\hline
  
\end{tabular}
\end{center}
\end{table*}

\begin{figure}[h]
\centering
 \includegraphics[width=0.5\textwidth]{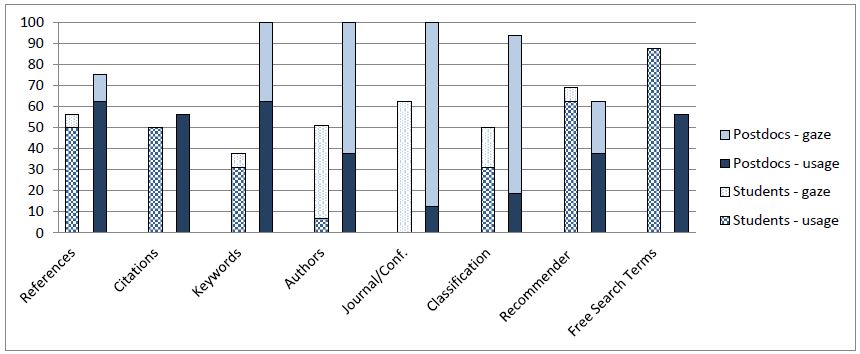}
\caption{Percentage of usage frequency and gaze data for the groups of students and postdocs.} 
\label{usagefreqvsgaze}       
\end{figure}

Besides statistically significant differences we found a general tendency towards a more intense focus on aoi by the group of postdoctoral researchers. With the exception of  citations and keywords, participants having a Ph.D. spent more time on these aoi and fixated these more often in comparison to the students.  
This tendency is displayed in more detail in Figure \ref{usagefreqvsgaze} in which the percentage of participants who \textit{gazed} at a particular element and those who \textit{used} such an element is shown. While the gaze data provided by the aoi is limited to the seed document in Figure \ref{usagefreqvsgaze} we summed up throughout the whole session. Differences between the two groups of participants become most evident for the meta data elements, the recommender and free  search term queries. Even though they did not use it, almost all postdoctoral researchers looked at all metadata, in particular keywords, authors, the journal or conference and the classification term, while only 37.5\% to 62.5\% of the students looked at them during their sessions. In turn, 67.5\% of the students and 62.5\% of the postdocs looked at the recommended articles, but while almost all of the students decided to follow them, only 37.5\% of the postdocs clicked on a recommended article. Another prominent difference between the two groups is that with 87.5\% a great amount of students conducted queries using free search terms, whereas only 57.5\% of the postdocs did so. Regarding the references and citations the gaze data and usage of the two groups differed only minimal.

\subsubsection{Stratagem usage pattern}

\begin{figure}[htbp]
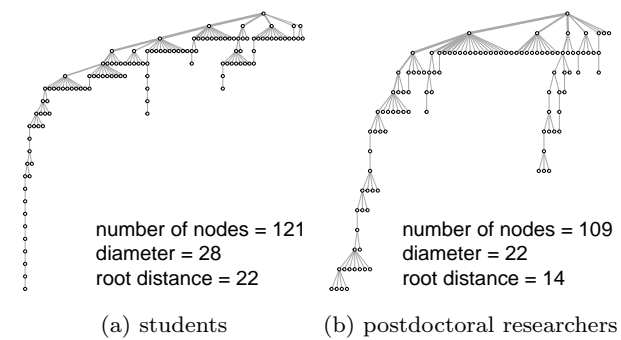

	\subfloat[students]{\includegraphics[width=0.48\linewidth]{students_graph.pdf}}
	\subfloat[postdoctoral researchers]{\includegraphics[width=0.48\linewidth]{phds_graph.pdf}}
	\caption{Combined session trees for (a) students and (b) postdoctoral researchers.} 
	\label{fig:jun_sen_graphs}
\end{figure}

Figure \ref{fig:jun_sen_graphs} shows the combined trees for the two groups students and postdoctoral researcher. At first glance, we can observe that the combined tree of the students group has a very long trail and that there is more interaction in level 4 to 6 than in the postdocs' tree. In contrast, the longer trail in the postdocs' tree has a broader activity, where multiple actions are conducted on one level. For further investigation of the combined session trees of our two groups, we create threshold graphs in the same manner we have done for the combined tree of all participants' sessions.

\begin{figure}[htbp]
	\subfloat[students]{\includegraphics[width=0.48\linewidth]{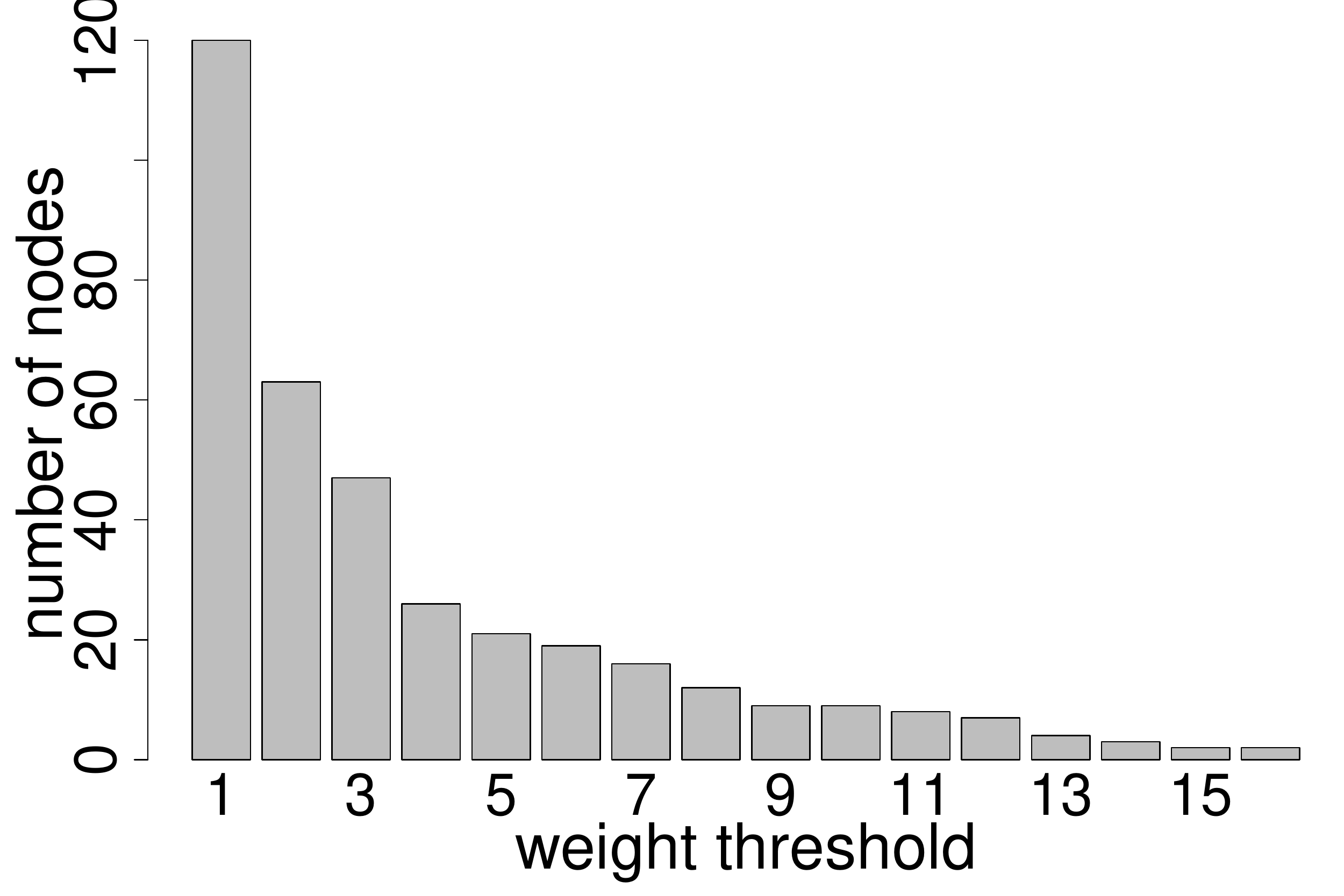}}
	\subfloat[postdoctoral researchers]{\includegraphics[width=0.48\linewidth]{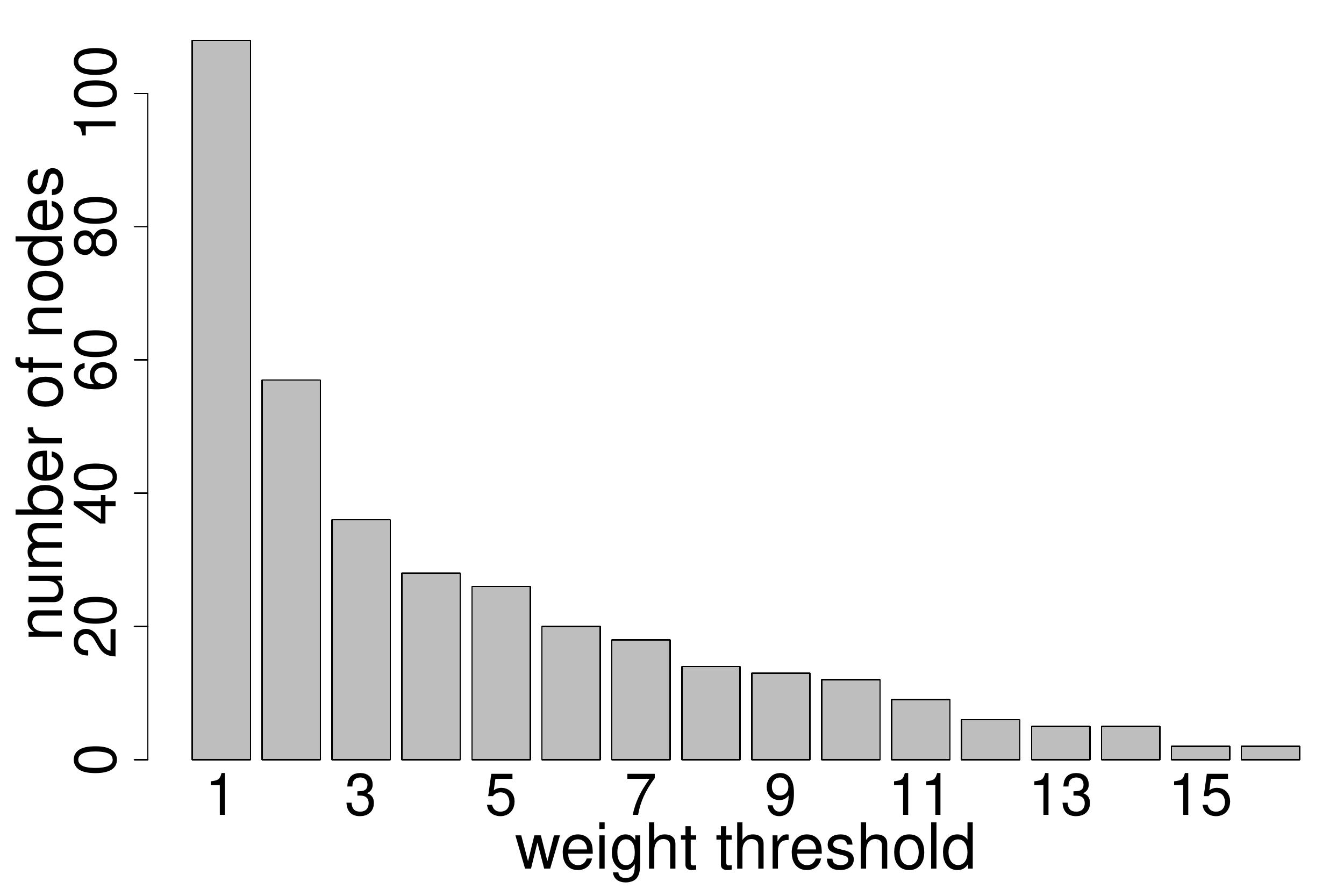}}
	\caption{Number of nodes left in the combined session trees of the students and the postdoctoral researchers group after removing all edges with a weight below a given threshold value.} 
	\label{fig:jun_sen_dist}
\end{figure}

Figure \ref{fig:jun_sen_dist} shows the threshold chart for the combined graphs of students and postdoctoral researchers. In the distribution for the students (cf. Figure \ref{fig:jun_sen_dist}a) we can see a small plateau for a threshold of 5. For postdoctoral researchers there is no clear plateau. Also, the trend for the students seems to be quite stable for higher thresholds. We compared all subtrees and decided to include the resulting trees for the three thresholds of 5, 9 and 12 for our analysis.

\begin{figure}[htbp]
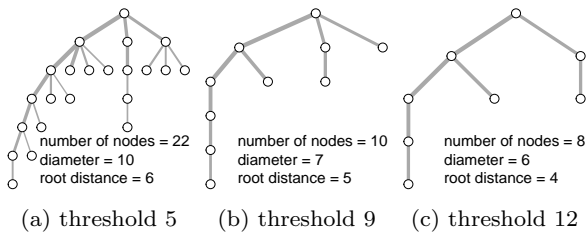

	\subfloat[threshold 5]{\includegraphics[width=0.31\linewidth]{students_graph_threshold_5.pdf}}
	\subfloat[threshold 9]{\includegraphics[width=0.31\linewidth]{students_graph_threshold_9.pdf}}
	\subfloat[threshold 12]{\includegraphics[width=0.31\linewidth]{students_graph_threshold_12.pdf}}
	\caption{Subtrees of the combined tree of students (cf. Figure \ref{fig:jun_sen_graphs}b).} 
	\label{fig:jun_thresholds_trees}
\end{figure}

\begin{figure}[htbp]
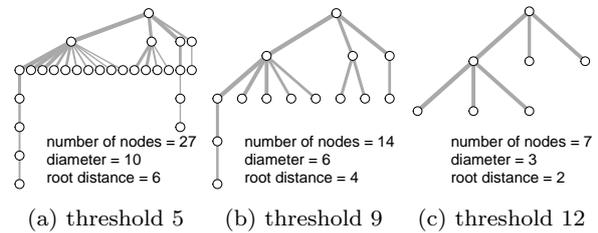

	\subfloat[threshold 5]{\includegraphics[width=0.31\linewidth]{phds_graph_threshold_5.pdf}}
	\subfloat[threshold 9]{\includegraphics[width=0.31\linewidth]{phds_graph_threshold_9.pdf}}
	\subfloat[threshold 12]{\includegraphics[width=0.31\linewidth]{phds_graph_threshold_12.pdf}}
	\caption{Subtrees of the combined tree of postdoctoral researchers (cf. Figure \ref{fig:jun_sen_graphs}a).} 
	\label{fig:sen_threshold_trees}
\end{figure}

Figure \ref{fig:jun_thresholds_trees} and \ref{fig:sen_threshold_trees} show the subtrees for both the students and the postdoctoral researchers group, for the thresholds of 5, 9, and 12. We can observe important differences between the two groups regarding a threshold of 5 (cf. Figure \ref{fig:jun_thresholds_trees}a and \ref{fig:sen_threshold_trees}a). For postdoctoral researchers the tree is very dense on the second level. For the students, there are more nodes that go deeper into the tree. Although diameter and root distance are equal, we can see that the number of nodes is higher for postdoctoral researchers, so there is more overall activity. Comparing the next pair for threshold 9 (cf. Figure \ref{fig:jun_thresholds_trees}b and \ref{fig:sen_threshold_trees}b), which represent at least the half of each group, we can see a similar relation. Postdoctoral researcher have slightly more activity and investigate broader than students. When looking at a threshold of 12 (cf. Figure \ref{fig:jun_thresholds_trees}c and \ref{fig:sen_threshold_trees}c), this observation can be seen even clearer. This threshold represents $3/4$ of the sessions. This subtree for the students has a higher root distance and is less broad. The postdocs' tree in contrast has more activity, which is intense on level 2 and 3.

%% file: discussion.tex
\section{Discussion}\label{discussionchap}
In the following section we discuss the results with respect to the research questions defined in Section \ref{RQINTRO}. \\\newline
\textbf{RQ 1: What are the most frequently applied stratagems in exploratory search in a state-of-the-art DL? How is the usage of stratagems in comparison to other search activities like the usage of recommendations and formulating queries?}

In total 137 search activities on the stratagem level  (cf . Table \ref{stratagemusagefrequency})
were performed of which keywords (50), references (27), citations (26),  and author information (25) were most frequently used. A rather low usage was found for the journal run (4) and classifications (5). One explanation for the little usage of a journal run was provided by five participants which stated that they did not perform a journal run because the regarding journal was too general for the task. Other reasons to reject a certain stratagem was that the use would have cost them too much time. 
Other search activities like queries and the usage of recommendations were utilised 111 times (cf . Table \ref{stratagemusagefrequency}). 
Starting from the seed document the participants most frequently utilised citations and keywords (cf . Table \ref{usageperwindow}). With increasing session steps the participants more frequently used queries and recommendations while the usage of citations and references decreased. This may, however, be due to missing citation and reference information in the documents discovered by the participants. With increasing session steps the percentaged usage of search activities varies only marginally. As mentioned by eight participants the continuous usage of a stratagem is influenced by the success of their previous search. One of the participants explicitly stated in the post-questionnaire to avoid mixing of search strategies. \\\newline
\textbf{RQ 2: Is there a common search pattern shared by the majority of the participants?} 

By creating combined session trees and deriving \linebreak subtrees from those, we were able to identify common patterns. Starting at the given root document the majority of participants conducted at least three individual stratagems. This shows that the participants did in fact follow multiple paths from the root document, which means, that they examined different paths to find related publications. In addition, at least one stratagem lead to more intense activity and was the starting point for further stratagems usage. This implies, that upon finding relevant information, the participants did investigate more deeply into this direction. So to say, they explore promising paths. However, it seems that they do not follow multiple paths deeply in parallel. Unfortunately, it is not clear, whether more intense exploration of parallel trails would have occurred when the participants would have had more time to search. Ten minutes might just have been too short to follow multiple paths. \\\newline
\textbf{RQ 3: Are there any differences in the usage of stratagems between students and postdoctoral researchers?} 

To look for significant differences between the two groups of participants we compared their search behaviour on three levels: search activity usage, eye tracking data, search patterns. 
\subsubsection*{Search activity usage}
Postdoctoral researchers performed 128 search activities of which keywords (35) and queries (33) were the most frequent (cf . Table \ref{tab:multicol}).
Students on the other hand performed less search activities (120) most frequently on the level of queries (36) and recommendations (24), while a journal run was utilised by none of the students. Most noticeable differences were found for the author run which was utilised nine times by the students and sixteen times by postdoctoral researchers. As stated in the post-questionnaire prior knowledge seems to play a vital role when deciding for a certain search activity. If the participants surmise or know that an author focuses on the topic of interest, they will be likely to perform an author run.
Although a difference in search behaviour between the two groups of participants was found, none of the results were statistically significant.   
\subsubsection*{Eye tracking data}
The gaze data provided by the eye tracking device \linebreak  showed significant differences between the two groups of participants for the seed document (cf. Table \ref{sigusagefrequencyeytracker}). Postdoctoral researchers spent more time inspecting the  journal information of the seed document. The group of students spent significantly (p=0.01) less time on this aoi (0.6 seconds) than the senior researchers (2.8 seconds). Likewise, the number of fixations on the journal was significantly lower (p=0.02) for the students (m=2.55 compared to m=8.33).
Both groups of participants spent the most time on inspecting the list of references. The mean dwell time was 69 seconds for postdoctoral researchers and 38 seconds for the group of students. 
Overall, postdoctoral researchers spent more time on nearly all metadata information of the seed document. The only exceptions are citations which were more intensively inspected by the group of students. We can assume that certain metadata information are more valuable to experienced researchers and thus, inspected more intensively (cf. \cite{Mayr2016}). However, the decision on utilising a stratagem strongly depends on the underlying content.  
\subsubsection*{Search pattern}
Regarding the search pattern, we can see differences between the two groups. Students
follow longer consecutive paths but inspect the result lists with less activity. Senior researchers start their search almost exclusively from the seed document and inspect the results more intensely, but follow longer trails less often. In alignment with \cite{aula2005eye}, \cite{buscher2012largescale} and \cite{white2007investigating}, the pattern seen for students could be categorised as economic (explorative) behaviour and the pattern of the senior researchers as exhaustive (navigational) behaviour. In these terms, \linebreak longer trails (depth) in our session trees indicate economic behaviour and many nodes on the same level (breadth) indicate exhaustive behaviour. However, we also see that there is economic and exhaustive behaviour in both groups. It is unclear in how far another group separation would create a more distinct picture regarding the search pattern.

\subsection{Implications for the design of Digital Libraries}
The results of the present study can have several implications for the design of DLs. 
\begin{itemize}
\item Linking content: The frequent usage of stratagems suggests that the participants consider structured information as potentially useful when exploring a DL. This could be further supported by interconnecting related content in a more intense way.  Possible interconnections could be revealed by applying bibliometric features like bibliographic coupling or author networks \cite{Carevic2015,Mutschke2015}.   
\item Extracting references and citations: A lot of research today is invested in extracting references and linking those to related documents. The present study shows that those information are frequently utilised and considered as very useful by the participants. However, the number of interconnected documents via citations and references is still comparably low. Further research should be invested to increase this interconnection to help users in exploring DL. 
\item Topical narrowing: The post-questionnaire provided us with valuable insights on the reasons for utilising or rejecting a certain search activity. The reason for the very limited usage of the journal run for example was a topical broadness which is not focused enough to find something related. DL could be designed to narrow down the content along the users' task by contextualizing search activities for example by aligning content along previous actions like the entered query term or the inspected document \cite{Carevic2015}. 
\item User guidance: Users of DL could further be supported in exploring DL by suggesting certain search activities that are likely to be useful for a given search situation. Comparable approaches can be found in the literature and have already been briefly discussed in \citep{carevic2016survey}. In \cite{Brajnik2002} a coaching approach is developed that provides the user with strategic help on potentially useful search activities that were derived from \cite{bates1990should} using a rule-based mechanism. A similar approach was presented in \cite{Kriewel/Fuhr:07,Kriewel/Fuhr:10} where an \em Adaptive Support for Digital Libraries \em was developed that covers sixteen predefined search activity suggestions. A more user oriented approach would be to identify search activities that expert users perform and use their search behaviour as a search strategy suggestion. This was also the main observation in \cite{fields2005designing}. 
\item Common search patterns: We were able to find search patterns that align with the user classifications defined in \cite{aula2005eye}, \cite{buscher2012largescale} and \cite{white2007investigating}. This could be exploited to generate behaviour specific recommendations and user interfaces as also proposed in \cite{white2007investigating}. Moreover, we were able to merge multiple patterns into a combined pattern. This approach could be used to generate a clustering method that is based on the usage pattern itself. This could lead to the identification of further user groups.

\end{itemize}

\subsection{Strengths and Limitations of the User Study}
Although it had thoroughly been pretested, the design of the user study showed some limitations.
The qualitative feedback of the participants indicates that ten minutes were too short for the task given in our scenario. Apparently, more time or even no time constraints at all might have been more appropriate for the given task and might have given more comprehensive insights on the order in which the participants use stratagems. 

Furthermore, it turned out that some participants were not familiar with Sowiport and the opportunities it offers, for which reason they lost some time during the task. A tutorial for Sowiport, e.g. a short demonstration or explanation followed by a few minutes for getting to know the digital library before the scenario was handed out would have increased the precision of the results. However, the decision not to introduce Sowiport made sure that the participants were not biased when focusing on certain areas of the interface. Furthermore, the demographics of the participants (see Section \ref{demographics}) showed that students are more likely to use Sowiport (25\% of the participants) compared to postdoctoral researchers (only 6\% of the participants). To validate whether this has an impact on our results we compared the user journey from the group of frequent Sowiport users (participants that used Sowiport often or very often) with non-frequent users. A preliminary examination of these showed no significant difference in their search behaviour. We therefor assume that the higher usage of Sowiport did not have an effect on the outcome of our results.

One could also assume that the seed document for the task was not perfectly chosen because of its very general scope. Though the journal led to almost no participant clicking and therefore influenced the user's behaviour. A different seed document with a more specialised journal would have influenced the participants' behaviour in the opposite way. 

Counting stratagem usage unveiled several problems. The counting is influenced by the circumstance that not all documents in Sowiport are provided with their references and citations. Also, in some cases participants forgot that they had already used a stratagem and used the same stratagem once again during the search task which leads to the question if these cases are counted as one or two stratagem usages. For this study we decided to exclude the multiple usage of search activities for a certain document. 

Similarly, the counting of the found relevant documents per stratagem proved to be difficult: Some stratagems open up a whole list of results for the user (e.g. the click on an author or keyword initiates a new search), while others lead to only one particular document (like the click on a reference or the click on a recommended document). In order to address this problem, we defined opening the reference or citation list in Sowiport, instead of clicking on a reference link, as one stratagem use, but with the lists containing different numbers of results, there remain limitations of their comparability. In addition, not all references and citations of a document in Sowiport are clickable and therefore influence the number of documents the participants were able to collect. Also, in several cases, participants used combinations of stratagems, for example by searching for an author and a keyword, or mixed them up, e.g. by conducting a free search with a keyword, which leads to further problems of relating a relevant document to a stratagem that has been used.

On the other hand, by using qualitative questions the user study enabled us to get an insight on the reasons why users decide to use a stratagem or not. The thorough analysis of the eye tracking data, including the selection of each participant's stable eye and the elimination of overlaying dynamic elements, is clearly a strength of our user study. Comparing the actual usage of search activities and gaze data provided by an eye tracking device we acquired an exhaustive view on search activities on the stratagem level. 
For this study we utilised an existing DL what makes the present study reasonably close to a real search task and search behaviour. 
As another benefit the participants were allowed to navigate in Sowiport without any restriction. By recruiting the participants from two levels of experience we were able to evaluate the search behaviour for two distinct groups. 

Analysing the search behaviour based on a tree based representation, has given us some valuable hints to potential user groups and commonalities. However, this technique should be improved in the future. It should be investigated what information is lost by ignoring the types actions conducted and in how far this approach could be extended, to be able to include different types activities.

%% file: conclusion.tex
\section{Conclusion}
In this paper we have presented the results of a user study on exploratory search activities in Sowiport, a DL for the social sciences. We recruited 32 participants, 16 students and 16 postdoctoral researchers all from the field of social sciences. In the first part of this paper we investigate the general usage of stratagems when looking for related work. The results of our study show that search activities on the stratagem level are frequently utilised by both groups of participants. The most frequently utilised search activities on the stratagem level were keywords search, following citations and references, and an author run. By comparing the actual usage of certain search activities and gaze data we acquired an exhaustive view on search activities on the stratagem level. To the best of our knowledge the usage of an eye tracking device to investigate and compare stratagems in DLs is a novelty and provided us with valuable insights. We showed that the participants frequently assess and use stratagems for further exploration. A stronger interconnection between documents in DLs could further support this process.   \\
Using a graph representation of the participants' search sessions and combining multiple sessions we were able to find common search patterns that the majority of our participants shared. We could see that the two kinds of behaviour exhaustive and economic could be identified using our graph representation. \\
By comparing the two groups of participants we found significant differences in the gaze data for the journal information which was fixated (number of fixations and dwell time) more intensely by postdoctoral researchers. The analysis of the gaze data showed that postdoctoral researchers in general focus on stratagems more intensely though not statistically significant. This may be due to the relatively small sample size. It would be interesting to further investigate this on a larger scale. 

The present work is part of an ongoing investigation of exploratory search activities in DLs. Future work will be a large scale investigations of stratagems by conducting a transaction log study in Sowiport.

%% file: acknowledgement.tex
\section{Acknowledgement}
This work was partly funded by DFG, grant no. MA 3964/5-1; the AMUR project at GESIS. We thank all participants of our user study, Dagmar Kern and the Sowiport \cite{Hienert2015} team at GESIS for supporting our study. 